\documentclass[journal=jpclcd,manuscript=letter,layout=twocolumn,10pt]{achemso}

\usepackage{chemformula} 
\usepackage[T1]{fontenc} 
\usepackage[version=3]{mhchem}
\usepackage{enumerate}
\usepackage{amsmath}
\usepackage{mathrsfs}
\usepackage{amssymb}
\usepackage{subfig}
\usepackage{stackrel}
\usepackage{amscd}
\usepackage{physics}
\usepackage{natbib}
\usepackage{array}
\usepackage{listings}
\usepackage{lmodern}
\usepackage{mathpazo}
\usepackage{microtype}
\usepackage{float}
\usepackage{caption}
\usepackage{setspace}
\usepackage{xkeyval}
\usepackage{color}
\usepackage{siunitx}
\usepackage[unicode=true,bookmarks=true,bookmarksnumbered=false,bookmarksopen=false,breaklinks=false,pdfborder={0 0 0},backref=true,colorlinks=true, citecolor=blue, urlcolor=blue]{hyperref}

\author{Jerryman A. Gyamfi}
\email{jerrymanappiahene.gyamfi@kuleuven.be}
\author{Thomas-C. Jagau}
\email{thomas.jagau@kuleuven.be}
\affiliation[KUL]{Department of Chemistry, KU Leuven, Celestijnenlaan 200F, B-3001 Leuven, Belgium}

\title[CAP-AIMD of temporary anions]{\textit{Ab initio} molecular dynamics of temporary anions 
using complex absorbing potentials}

\abbreviations{CAP, AIMD, CAP-AIMD}

\begin{document}

\begin{tocentry}
\includegraphics[scale=1.0]{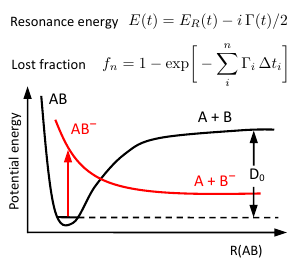}
\end{tocentry}


\begin{abstract}
Dissociative electron attachment, that is, the cleavage of chemical bonds induced by low-energy 
electrons, is difficult to model with standard quantum-chemical methods because the involved 
anions are not bound but subject to autodetachment. We present here a new computational 
development for simulating the dynamics of temporary anions on complex-valued potential 
energy surfaces. The imaginary part of these surfaces describes electron loss, whereas the 
gradient of the real part represents the force on the nuclei. In our method, the forces are 
computed analytically based on Hartree-Fock theory with a complex absorbing potential. 
\textit{Ab initio} molecular dynamics simulations for the temporary anions of dinitrogen, ethylene, 
chloroethane, and the five mono- to tetrachlorinated ethylenes show qualitative agreement 
with experiments and offer mechanistic insights into dissociative electron attachments. The 
results also demonstrate how our method evenhandedly deals with molecules that may undergo 
dissociation upon electron attachment and those which only undergo autodetachment.
\end{abstract} 


Temporary anions (TAs) \cite{herbert15,ingolfsson19} are formed when a neutral molecule captures 
a free electron, most often one with low energy below 10 eV. Albeit their lifetime is in the range of 
femto- to milliseconds, TAs play a central role in our understanding of important chemical processes: 
these include DNA lesions in living cells upon exposure to ionizing radiation \cite{boudaiffa00,simons07}, 
the formation of radicals and anions in the Earth's atmosphere \cite{rees89,rankovic20} and plasma 
etching in the semiconductor industry \cite{stoffels01}. TAs are also currently reshaping our understanding 
of the physics and chemistry in interstellar and circumstellar environments \cite{millar17}.

Studying TAs by means of electronic-structure methods designed for bound states, that is, states 
related to the discrete spectrum of the electronic Hamiltonian, is difficult, and in many cases a 
hopeless venture. To begin with, temporary anions are electronic resonances, i.e., metastable 
states embedded in the continuum of the electronic Hamiltonian\cite{moiseyev11,jagau22,jagau17}. 
Secondly, the formation of a TA may lead to fragmentation of the molecule, which is commonly 
referred to as dissociative electron attachment (DEA). If DEA is possible, it competes against 
autodetachment, which makes the computational modeling and prediction of the dissociation 
channels an intricate problem in its own right \cite{simons81,asgeirsson16,fabrikant17,ingolfsson19}. 

Various methods have been proposed to address the difficulties associated with the modeling of 
TAs and DEA; recent overviews are available in Refs. \citenum{jagau22,masin20}. A group of 
methods combine scattering theory and operator projection techniques\cite{feshbach58,taylor06,
goldberger04}, for example, the approach by Domcke\cite{domcke91} or the Schwinger multichannel 
approach by Takatsuka and McKoy\cite{takatsuka84}. In these methods, an explicit treatment of the 
continuum is necessary. Recently, Kossoski and co-workers reported an implementation of \textit{ab initio} molecular 
dynamics (AIMD) for studying TAs.\cite{kossoski19} They used bound-state methods to compute the 
resonance energies and determined the corresponding lifetime using the Schwinger multichannel 
method. 

In this letter, we present a new computational development, dubbed CAP-AIMD, which integrates complex 
absorbing potentials (CAPs) \cite{riss93} into AIMD simulations\cite{marx09}. In our approach, we avoid scattering 
calculations and obtain the energy and lifetime of a TA as well as the forces acting on the nuclei 
in a single computation at each time step. As initial applications, we report CAP-AIMD results for 
the temporary anions of dinitrogen, ethylene, chloroethylene, the three isomers of dichloroethylene, 
trichloroethylene, tetrachloroethylene, and chloroethane. It is well established that the chlorinated 
compounds all undergo DEA while dinitrogen and ethylene do not \cite{fabrikant17}; the CAP-AIMD 
approach handles both cases evenhandedly.  


The fundamental difference between conventional AIMD and CAP-AIMD is that in the latter, the nuclei 
evolve on a complex potential energy surface (CPES). The CPES is obtained from a non-Hermitian 
Hamiltonian with complex eigenvalues
\begin{equation} \label{eq:Siegert_energy}
E = E_r - i\Gamma/2 \ ,
\end{equation}
called Siegert energies \cite{siegert39}. Here, $E_r$ is the energy of a resonance state and $\Gamma$ 
its decay width, which is inversely proportional to the lifetime $\tau$. For states, which are electronically 
bound, $\Gamma=0$. 

In our approach, the CPES is computed on the fly using the CAP method. The CAP Hamiltonian $H^\eta$ 
is obtained from the physical Hamiltonian $H$ as \cite{riss93,sommerfeld98,santra01,zuev14,dempwolff22,
gayvert22}
\begin{equation} \label{eq:H_eta}
H^\eta = H - i \eta W \ ,
\end{equation}
where $\eta$ is a real scalar parameter and $W$ is the CAP. We choose $W$ to be quadratic in the 
electronic coordinate. Mathematical and technical details of this CAP have been discussed elsewhere 
\cite{riss93,santra01,zuev14,jagau14}. The important thing to note here is that $W$ depends on a set 
of three parameters, $\{r^o_\alpha\}, \ \alpha = x,y,z$, which define its onset. Hence, the CAP is 
parameterized by the quartet $\{r^o_\alpha, \eta\}$, which needs to be optimized in order to minimize 
the perturbation of the resonance wave function. In principle, this needs to be redone at each time step 
as the nuclear configuration changes. We have found, however, that determining the CAP parameters 
for the initial structure of the TA and keeping them fixed during the AIMD simulation until the anion 
becomes bound produces meaningful results. When the anion is bound, $\eta$ is set to zero. Details 
on the procedure for determining the optimal quartet $\{r^o_{\alpha,opt}, \eta_{opt}\}$ are discussed 
in the Supporting Information. 

Although it is possible to generate the CPES using other complex-variable techniques \cite{moiseyev11,
jagau22}, we chose here the CAP method because of the availability of analytic gradients \cite{benda17,
benda18}. Similarly, while one can compute CPES at a higher level of electronic-structure theory, we 
limit ourselves to Hartree-Fock (HF) theory here because of the computational demands of AIMD 
simulations. The low level of theory notwithstanding, our CAP-AIMD results show qualitative agreement 
with experiments and offer insights into the connection between the initial TA and the final DEA products. 
Questions like, for example, how the molecular orbitals (MOs) of the TA evolve in time may also be answered. 

We propagate the nuclei on the CPES according to the classical Hamilton equations
\begin{align}
\frac{d{\vb{R}}_k}{dt} & = \frac{\vb{P}_k}{M_k} \label{eq:coordinate_nucleus} \\
\frac{d{\vb{P}}_k}{dt} & = \vb{F}_k = -\grad_k \left(\Re E + V_{nuc-nuc}  \right) \label{eq:force_nucleus}
\end{align}
where $\vb{R}_k$, $\vb{P}_k$ and $M_k$ are the coordinate, momentum vector and mass of the 
$k-$th nucleus, and $V_{nuc-nuc}$ is the nuclear repulsion energy. Note that the force $\vb{F}_k$ 
depends on \textit{only} the real part of $E$. Indeed, if the nuclei are treated classical and nuclear 
quantum effects are ignored, the force $\vb{F}_k$ is simply proportional to the gradient of the real part 
of the CPES \cite{moiseyev17}. 

After running a CAP-AIMD simulation, one obtains a collection $\{E_n\}$ of complex energies, one for 
each time step $n$. The imaginary part of $E_n$ relates to the resonance width $\Gamma_n$ through 
Eq. \eqref{eq:Siegert_energy}. The $\{\Gamma_n\}$ profiles can be used to distinguish TAs, which 
undergo only autodetachment from those which undergo dissociation. In the first case, $\Gamma$ 
oscillates with time, while in the second case it drops to zero at a critical point and stays so indefinitely. 

Furthermore, one can estimate from $\Gamma_n$ the probability $P_{n+1}$ that the TA survives 
autodetachment between steps $t_{n}$ and $t_{n+1}$. Taking $\Gamma_n$ as the width in the 
interval $t_{n}\leq t < t_{n+1}$ leads to the expression
\begin{equation}
P_{n+1} =  e^{- \Gamma_n \cdot \Delta t_{n+1}} = e^{- \Delta t_{n+1}/\tau_n  }\ ,
\end{equation}
where $\Delta t_{n+1} = t_{n+1} - t_{n}$. This probability may be employed to estimate the 'lost 
fraction', $f_n$, that is, the fraction of an ensemble of TAs, all characterized by exactly the 
same initial conditions, statistically expected to undergo autodetachment \emph{by} the time 
step $t_n$. In the Supporting Information, we show that
\begin{equation} \label{eq:F_n_final}
f_n   = 1 - \prod^{n}_{i=1} P_i  = 1 - e^{-  \sum^{n}_{i=1} \Gamma_{i-1} \cdot \Delta t_i} \ .
\end{equation}
It must be noted that the summation in the exponent in Eq. \eqref{eq:F_n_final} is an approximation 
to an integral of $\Gamma$ over time in the continuous time limit. Hence, for a good estimate of 
$f_n$, one needs to choose the time step $\Delta t$ such that $\Gamma_0 \Delta t < 1 $, where 
$\Gamma_0$ is the resonance width at the TA's initial geometry\cite{note:Delta_t}.

For meaningful results, it is paramount to conduct the CAP-AIMD simulation with the desired HF 
state, meaning the self-consistent field (SCF) solution that corresponds to the resonance and not 
some discretized continuum state. Indeed, as we show in the Supporting Information, the dynamics 
of pseudocontinuum states are fundamentally different from those of a resonance. We use the 
following procedure to find the right SCF solution in the continuum: i) at time step $n=0$, we build 
the SCF guess from the core Hamiltonian, then, ii) at any subsequent time step $n>0$, we use the 
SCF solution found at the $(n-1)$-th step as guess. As indicated above, the CAP is turned off as 
soon as the anion is bound. In order to determine this, we use Koopmans' theorem and consider 
the real part of the energy of the highest occupied MO (HOMO): If it is negative, we 
consider the anion to be bound. 

To judge how reliable a CAP-AIMD simulation is, we compute in every time step $n$ a deperturbative 
correction to the complex energy according to \cite{riss93, jagau14}
\begin{equation}
\widetilde{E}_n  = E_n - \eta \frac{dE_n}{d\eta} = E_n + i \eta \langle W \rangle_n~. 
\end{equation}
For a well-represented resonance, $\widetilde{E}_n$ and $E_n$ deviate only little from each 
other, whereas this is not the case for pseudocontinuum states. Thus, in general, the closer $E_n$ 
and $\widetilde{E}_n$ are, the better the simulation. Note that the deperturbed lost fraction, 
$\widetilde{f}_n$, may also be computed by replacing $\Gamma_i$ in Eq. \eqref{eq:F_n_final} with 
$\widetilde{\Gamma}_i = -2 \Im \widetilde{E}_i$.

We implemented our method in the Q-Chem program \cite{qchem50} making use of the implementation 
of CAP-HF energies \cite{zuev14} and analytic gradients \cite{benda17,benda18}. The results reported 
below should be viewed as proof of concept and as illustration of the robustness of CAP-AIMD 
simulations. For these reasons, we performed no averaging over initial structures and always took 
the optimized geometry of the neutral molecule, determined at the HF level, as the initial geometry 
of the corresponding TA, unless stated otherwise. This is equivalent to a vertical electron attachment  
at the neutral equilibrium geometry and simulating the time evolution of the formed TA. The cc-pVTZ 
+ 3p basis set with the diffuse p-functions placed on all atoms except hydrogen was used for all 
simulations and the initial geometry optimizations. Only for tetrachloroethylene, we used aug-cc-pVDZ 
+ 3p instead. The number of trajectories for each TA discussed below ranges between 20 and 200; 
all ran in the microcanonical ensemble.


\begin{figure*}
\includegraphics[scale=0.54]{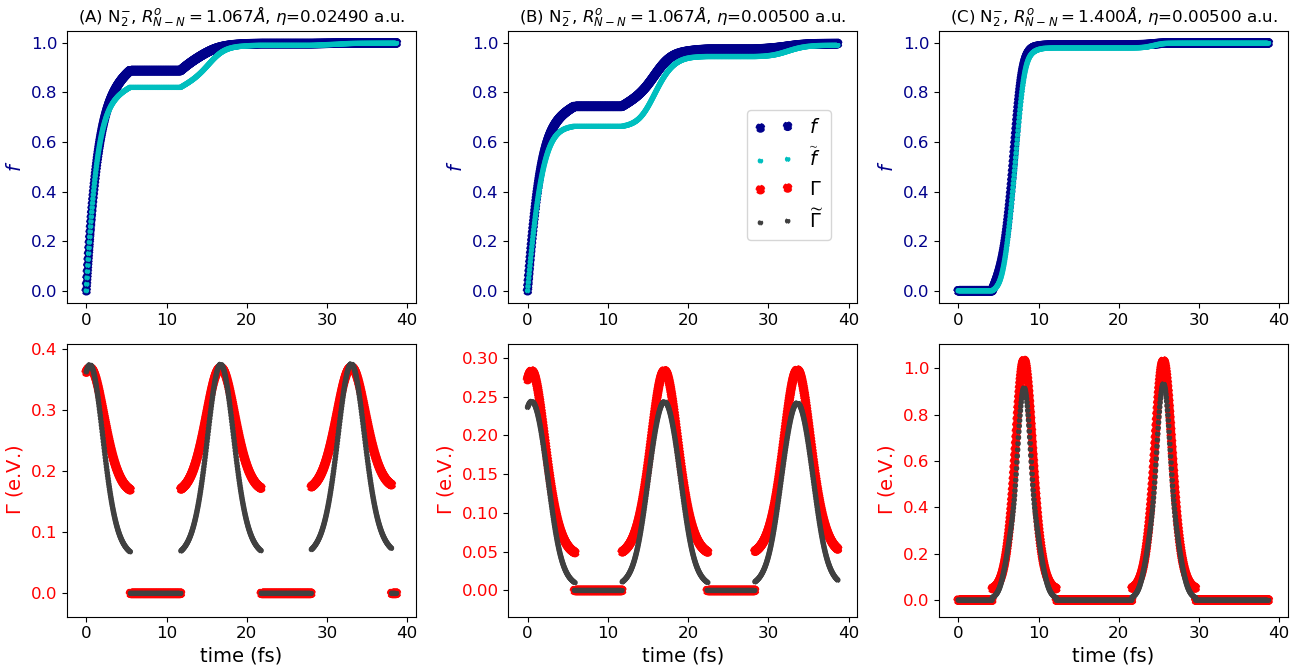} 
\caption{Lost fraction, $f$, and resonance width, $\Gamma$, as well as their deperturbed counterparts, 
$\widetilde{f}$ and $\widetilde{\Gamma}$, for randomly chosen CAP-AIMD trajectories of \ch{N2-} 
simulated under different conditions. The profiles are limited to the first 40 fs after vertical electron 
attachment but $\Gamma$ and $\widetilde{\Gamma}$ remain periodic in the entire duration of the 
simulations, which is $\sim 140 $ fs. Time step is 2 a.u. $\approx$ 0.05 fs.}
 \label{fgr:n2}
\end{figure*}

\begin{figure}[th]
\includegraphics[scale=0.29]{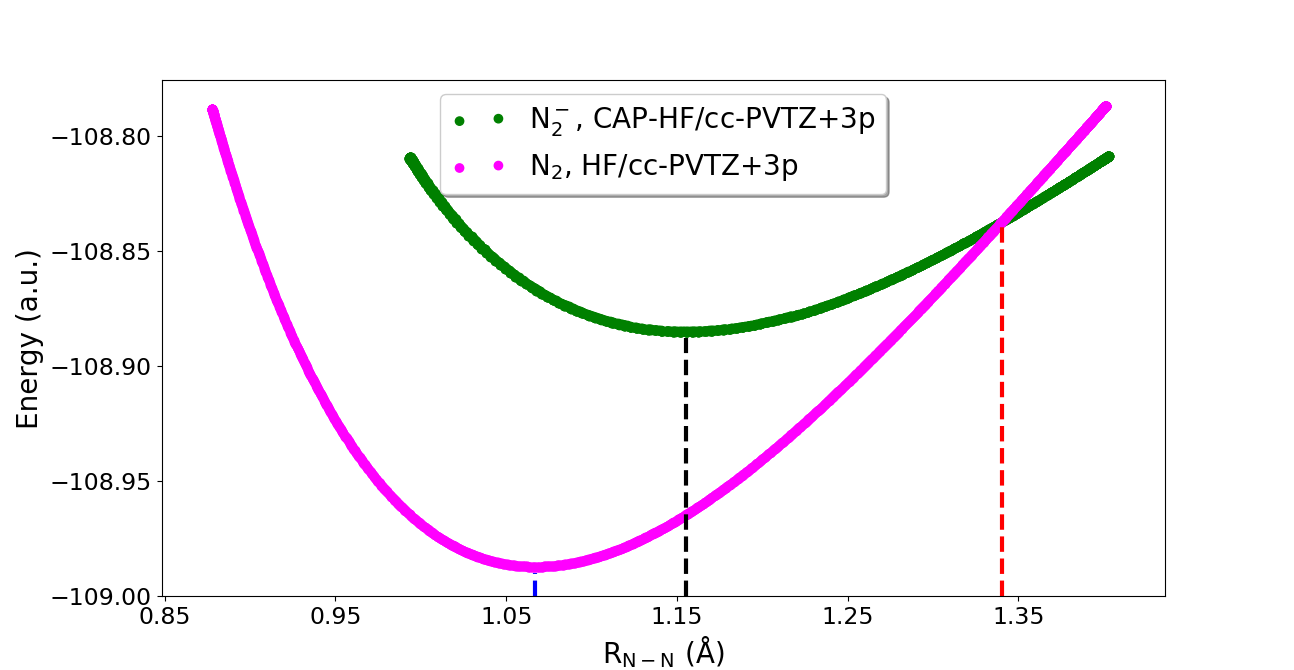} 
\caption{Real part of the CPES of \ch{N2-} (in green) and the PES of \ch{N2} (in magenta). Time step 
is 2 a.u. $\approx$ 0.05 fs, total number of time steps is 2000. The curve for \ch{N2} derives from a 
standard AIMD simulation, while that of \ch{N2-} is from a CAP-AIMD simulation.}
\label{fgr:N2PES}
\end{figure}

The $^2\Pi_g$ shape resonance of \ch{N2-} is an example of a TA which only undergoes autodetachment 
but does not induce fragmentation\cite{schultz73}. A similar case is the TA of ethylene, for which 
we report results in the Supporting Information. In Fig. \ref{fgr:n2}, we report resonance widths 
$\Gamma$ and lost fractions $f$ for \ch{N2-} from randomly chosen CAP-AIMD trajectories. In 
Fig. \ref{fgr:N2PES}, we show the real part of the CPES derived from the same simulations. Cases 
(A) and (B) in Fig. \ref{fgr:n2} refer to simulations started at the same initial N-N bond length of 
1.067 \AA\ with the same CAP box dimensions but different $\eta$ values: $\eta= 0.02490$ a.u., 
which is the optimal $\eta$ for the initial geometry, in case (A) and $\eta= 0.00500$ a.u. in case 
(B). In case (C), we started the simulation at an initial N-N bond length of $1.400$~\AA, where 
\ch{N2-} is bound (see Fig. \ref{fgr:N2PES}) and chose $\eta=0.00500$ a.u. for the unbound region.

The resonance width has a cyclic profile in all three cases in Fig. \ref{fgr:n2} because of the vibration of the molecule. We observe a direct 
correlation between $\Gamma$ and the N-N bond length. Segments of the $\Gamma$ profile in 
Fig. \ref{fgr:n2} where we see a decrease (increase) correspond to stretching (shortening) of the bond 
below $1.20$ \AA, whereas segments with $\Gamma = 0$ correspond to $R_{N-N} \gtrsim 1.20$ \AA. 
When $\Gamma$ is nonzero, we also see an increase in the lost fraction, whereas it remains constant 
while $\Gamma = 0$. This suggests that \ch{N2-} is unbound when $R_{N-N}\lesssim 1.20$ \AA\ and 
that the potential curves of \ch{N2} and \ch{N2-} cross at $R_{N-N}\approx 1.20$ \AA\ according to Koopman's theorem. But Fig. \ref{fgr:N2PES} shows that the potential curves cross at $R_{N-N} \approx 1.33$ \AA\ if independent HF calculations are performed for \ch{N2} and \ch{N2-}; 
in fact, it is known that methods that describe a TA and the neutral molecule with the same Hamiltonian 
yield a more consistent description \cite{jagau14b}. (See Supporting Information for a plot of the imaginary part of the CPES of \ch{N2-}.)

The similarity between cases (A) and (B), which differ only in the $\eta$ value, indicates some 
flexibility in choosing this parameter. Overall, we have better agreement between the uncorrected 
$\Gamma$ (in red) and the deperturbed $\widetilde{\Gamma}$ (in dark gray) in case (B). Also, the 
discontinuity at $\Gamma \to 0$ is smaller. However, there is good agreement between the uncorrected 
($f$) and corrected ($\widetilde{f}$) lost fractions in both cases (A) and (B). In case (C), where we start 
at $R_{N-N} = 1.40$ \AA, the lost fraction curve is different from (A) and (B); $\Gamma$ stays at zero 
for the first $\sim 4.2$ fs. In that initial interval, the N-N bond compresses to about $1.2$\AA. Thereafter, 
as the N-N bond continues to compress, the autodetachment process ensues and $\Gamma$ begins 
to increase. Because the nuclei gain more kinetic energy in case (C) than in (A) and (B), more compressed bond lengths and higher $\Gamma$ values are reached. Importantly, in all three cases in Fig. \ref{fgr:n2}, $f$ and $\widetilde{f}$ become $\sim 1$ 
within 40 fs meaning that it will be very rare to find an anion 40 fs after electron attachment, which is 
less than 3 vibration periods of neutral \ch{N2}. 

\begin{figure*}
\includegraphics[scale=0.54]{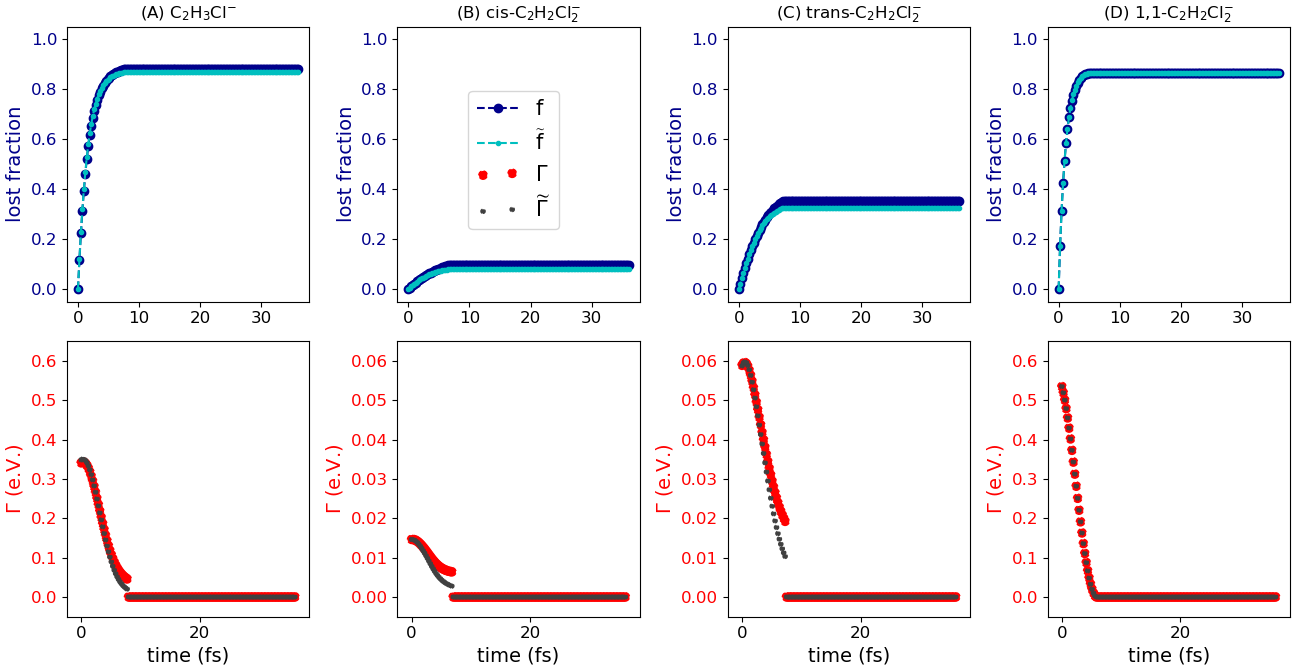} 
\caption{Lost fraction, $f$, and resonance width, $\Gamma$, as well as their deperturbed counterparts, 
$\widetilde{f}$ and $\widetilde{\Gamma}$, for randomly chosen CAP-AIMD trajectories of the anions 
of A) chloroethylene, B) \textit{cis-}, C) \textit{trans-} and D) \textit{1,1-} dichloroethylene. The profiles 
are limited to the first 40 fs.}
\label{fgr:cheth}
\end{figure*}

We now turn our attention to molecules which may undergo DEA. In Fig. \ref{fgr:cheth}, we show 
results from arbitrarily chosen trajectories for the anions of chloroethylene and the three isomers of 
dichloroethylene. Immediately, we see a stark difference with \ch{N2-}: $\Gamma$ does not oscillate 
but falls to zero in less than $\sim$10 fs, during which time we also see a steep rise and stabilization 
in the lost fraction profile. 

Dissociation is possible in all four cases of Fig. \ref{fgr:cheth}. The predominant dissociation channel 
we observe is the formation of \ch{Cl-} and an organic radical (see Supporting Information), which is 
in agreement with experiments \cite{vasilev08}. It is worth noting that all TAs become bound before 
dissociation takes place. The latter is marked by a stabilization of the anion's energy, which becomes 
discernible when one averages $\Re E$ over many trajectories (see Supporting Information). In our 
simulations, the C-Cl bond cleavage is complete 100-150 fs after electron attachment.

The important differences among the molecules presented in Fig. \ref{fgr:cheth} are the lost fraction 
profiles, which are determined by two factors (see Eq. \eqref{eq:F_n_final}): the resonance width 
and the time it takes a TA to become bound. Both factors relate to the shape of the CPES. In general, 
the closer the initial geometry of the TA is to a bound region on the CPES, the smaller the initial 
resonance width $\Gamma_0$. Also, the more the real part of the gradient at the initial geometry 
points towards the bound region, the lesser time it will take the TA to reach that region. 

Since the bound region is reached very quickly in all four simulations in Fig. \ref{fgr:cheth}, 
$\Gamma_0$ is the most important factor in explaining the lost fraction profiles. For the TAs of 
interest here, $\Gamma_0$ decreases in the following order: \ch{CCl2=CH2} > \ch{CHCl=CH2} 
> \textit{trans}-\ch{CHCl=CHCl} > \textit{cis}-\ch{CHCl=CHCl}. The equilibrium lost fractions 
$f_{\infty} \equiv \lim_{n \to \infty} f_n$ decreases indeed in the same order. The statistical 
analysis of our simulations (see Supporting Information) yielded for $\left< f_{\infty} \right > $ 
values of 0.93, 0.92, 0.33, and 0.10 for \ch{CCl2=CH2}, \ch{CHCl=CH2}, 
\textit{trans}-\ch{CHCl=CHCl}, and \textit{cis}-\ch{CHCl=CHCl}, respectively. 

These values allow us to estimate how effective DEA is: For example, for 1,1-dichloroethylene, an 
average of 7\% of the initially formed TAs live long enough for dissociation to happen. For \textit{trans}- 
and \textit{cis}-dichloroethylene, this average jumps to $67\%$ and $90\%$, respectively, resulting 
in a much larger \ch{Cl-} ion yield. This agrees well with experiments \cite{vasilev08}, where it was 
found that the DEA cross section for the \ch{Cl-} channel is highest for {cis}-\ch{CHCl=CHCl}, 
followed by \textit{trans}-\ch{CHCl=CHCl}, and \ch{CH2=CCl2}.

\begin{figure*}
\includegraphics[scale=0.5]{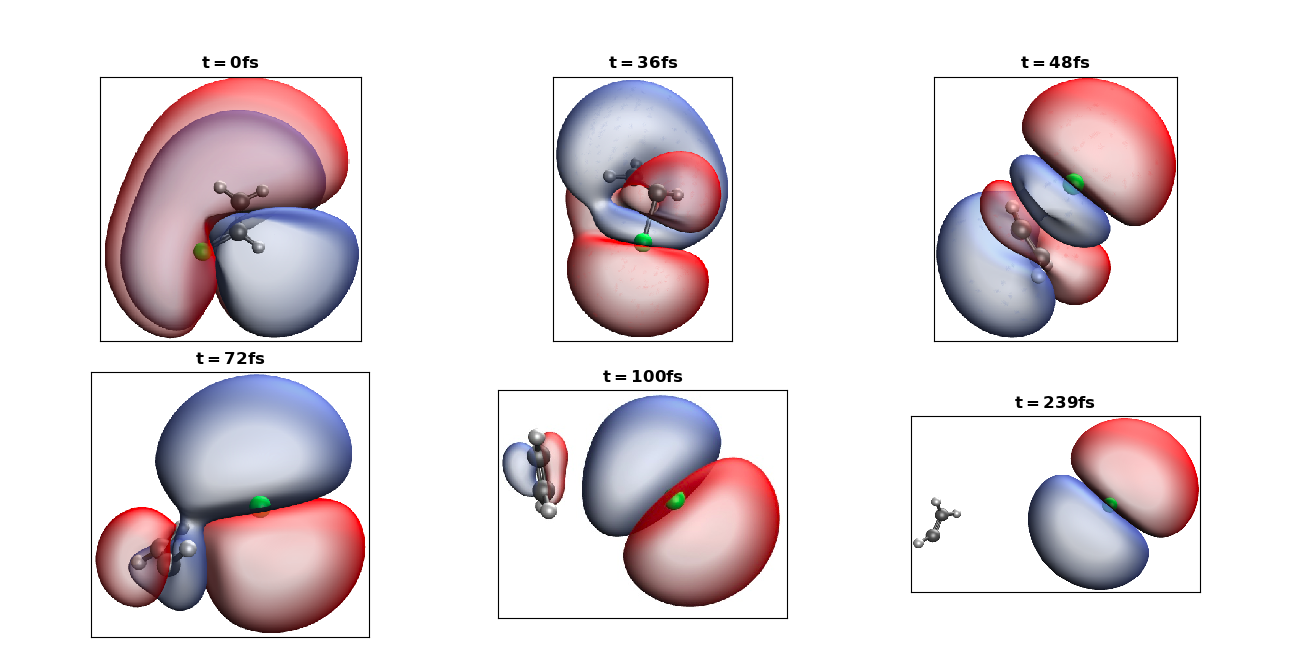} 
\caption{Time evolution of the HOMO of chloroethylene anion. Isovalue is 0.002.}
\label{fgr:mos}
\end{figure*}
 
The ethylene-derived TAs under discussion here are known to originate from the capture of an 
electron into the $\pi^*$ orbital of the C=C bond, giving rise to a $^2\Pi$ anion state \cite{burrow81,
olthoff85}. According to electron transmission spectroscopy (ETS), DEA proceeds, however, via 
a $^2 \Sigma$ state, which suggests a $^2\Pi \rightarrow\ ^2\Sigma$ transition mediated by an 
out-of-plane motion during the C-Cl bond elongation \cite{burrow81,olthoff85,clarke69,benda18b}. 
As discussed in the Supporting Information, our CAP-AIMD simulations confirm that. 

The $^2\Pi \rightarrow \ ^2\Sigma$ transition is also evident from the changing character of the 
HOMO along the trajectory; Fig. \ref{fgr:mos} illustrates this for the chloroethylene anion. These 
orbital plots were generated along the trajectory from Fig. \ref{fgr:cheth} (A). It is seen that the 
HOMO is of $\pi^*$ character at $t=0$, while we have mixed $\pi^*$-$\sigma^*$ character at 
48 fs. Notably, this plot is very similar to that of the coupled-cluster Dyson orbital computed 
at the minimum-energy crossing point between neutral and anionic chloroethylene \cite{benda18b}. 
Already at 72 fs, the HOMO is largely localized on the Cl atom and the Mulliken charge of this 
atom is already about $-0.9$ a.u.
 
 
In the Supporting Information, we also report CAP-AIMD results for the anions of trichloroethylene 
and tetrachloroethylene. Here too, we observe that the $^2\Pi \rightarrow\ ^2\Sigma$ transition is 
mediated by an out-of-plane motion. For trichloroethylene, we observe four dissociation channels: 
the three isomers of dichloroethylene together with \ch{Cl-}, and \ch{Cl + C2Cl- + HCl}.  All four 
channels have been observed in catalyzed reductive dechlorination \cite{neumann96, gantzer91,
glod97}; however, while formation of the \textit{trans}-dichloroethylene radical dominates in our 
simulations, the experiments all find that the \textit{cis}-dichloroethylene radical is dominant. This 
discrepancy may be explained by equilibration between the two radicals which is known to happen 
but it could also be related to stereo- and regioselectivity of the catalysts (e.g., vitamin B$_{12}$) 
used in the experiments \cite{nonnenberg02}. We note that our simulations do not take into account 
any environment and are performed at a low level of theory, i.e., HF. For tetrachloroethylene anion, 
we observe only the formation of trichloroethylene radical and \ch{Cl-}, which is consistent with 
experiments \cite{neumann96,gantzer91,glod97}. The experiments also find subsequent 
dechlorination of the trichloroethylene anion formed from the radical yielding the same DEA 
products discussed in the last paragraph. 

To test the CAP-AIMD method for DEA involving only a $\sigma^*$ resonance, we studied 
the TA of chloroethane. This is also reported in the Supporting Information; we find an equilibrium 
lost fraction very close to 1 indicating a small DEA cross-section consistent with experiment 
\cite{pearl94} and previous theoretical results \cite{kossoski19}.

It is worth noting that we did not observe dissociation without a CAP. Such simulations do 
not capture electron attachment but describe pseudocontinuum states. The notable exception 
is tetrachloroethylene, whose anion is almost bound at the neutral equilibrium structure making 
it possible to simulate DEA without a CAP. As shown in the Supporting Information, dissociation 
is also suppressed when too high or too low CAP strengths are used. Finally, we note that we observed in 
a small number of trajectories artificial imaginary energies long after the anion has become stable 
against electron loss. We ignored these artificial imaginary energies in the analysis of the results.

 
In summary, CAP-AIMD offers a robust method to simulate the dynamics of TAs, independent 
of whether they undergo dissociation or not. The nuclei are propagated on a complex potential 
energy surface, yielding a complex energy at each time step, where the imaginary part relates 
to the decay width of the TA. In addition, the fraction $f$ of TAs lost due to autodetachment can 
be obtained from CAP-AIMD simulations. The equilibrium value $\left< f_{\infty}\right>$ provides 
a measure for the efficiency of DEA. We conducted CAP-AIMD simulations for the anions of 
chloroethane and different chlorinated ethylenes; the observed trends in the DEA yield are 
consistent with experimental results. A remarkable commonality among these simulations is 
that all anions become stable towards autodetachment after less than 10 fs. We conclude by 
noting that an obvious challenge for future work is to improve the description of the electronic 
structure, that is, replacing HF by density functional theory in CAP-AIMD simulations. 


\begin{acknowledgement}
Funding from the European Research Council (ERC) under the European Union's Horizon 2020 
research and innovation program (Grant Agreement No. 851766) is gratefully acknowledged. 
Resources and services used in this work were partly provided by the VSC (Flemish Supercomputer 
Center), funded by the Research Foundation - Flanders (FWO) and the Flemish Government.
\end{acknowledgement}

\begin{suppinfo}
Derivation of Eq. \eqref{eq:F_n_final}; information on the CAP parameters used in our simulations 
and the procedure to determine them; more computational results for all anions studied (dinitrogen, 
ethylene, chloroethane, chloroethylene, \textit{cis}-dichloroethylene, \textit{trans}-dichloroethylene, 
\textit{1,1}-dichloroethylene, trichloroethylene, and tetrachloroethylene).
\end{suppinfo}

\bibliography{a_CAP_AIMD}

\end{document}


\section{Lost fraction formula derivation}
Following the article, let $P_{n+1}$ be the survival probability of the temporary anion between the 
${n}-$th and $(n+1)$-th steps of the CAP-HF AIMD simulation. Then, 
\begin{equation}
\begin{cases}
P_0     & = 1\\
P_{n+1} & = e^{- \Gamma_n \cdot \Delta t_{n+1}} \ , \qquad n \in \{ 0,1,2 \ldots \}\ ,
\end{cases}
\end{equation}
where $\Delta t_{n+1} \equiv t_{n+1} -t_{n}$, and $\Gamma_n \equiv -2\cdot \Im E_n$. $E_n$ is the 
complex SCF energy at the $n-$th AIMD step. $P_0$ defines the survival probability \emph{at} the 
initial instance $t_0$, and is set to be unity. The probability that we witness autodetachment between 
the $n-$th and $(n+1)-$th steps, after the temporary anion had survived down to the $n-$th step, is:
\begin{equation}
\begin{split}
A_{n+1} & = \left[\prod^{n}_{i=0} P_i\right] \cdot (1-P_{n+1}) \ \qquad n \in \{ 0,1,2 \ldots \} \ .
\end{split}
\end{equation}
We set $A_0 = 0$ since there is no autodetachment at the initial instant $t_0$.

Consider now an ensemble $N$ of identical temporary anions who share the same initial geometry and initial 
velocities. It is sufficient to run a single CAP-AIMD simulation for this particular ensemble to obtain 
the data set $\{\Gamma_n\}$, which informs us of the width of each member of the ensemble along 
that given trajectory. Without loss of generality, we may take $A_{n+1}$ to be equivalent to the 
fraction of the initial number $N$ of temporary anions we lose  between time steps $n$ and $n+1$. 
Thus, the sum
\begin{equation}
f_n \equiv A_0 + A_1+A_2+\ldots +A_n
\end{equation} 
is the total fraction lost between time steps $0$ and $n$ (included). 
That is, for $n\geq 1$,
\begin{equation} \label{eq:F_n}
f_n  = \sum^n_{m=1} A_m = \sum^n_{m=1} \left[\prod^{m-1}_{i=0} P_i\right] \cdot (1-P_m) \ ,
\end{equation}
which can be simplified to:
\begin{equation} \label{eq:F_n_final}
f_n   = 1 - \prod^{n}_{i=1} P_i  = 1 -e^{-1/\hbar \  \sum^{n}_{i=1} \Gamma_{i-1} \cdot \Delta t_i} \ .
\end{equation}

There are a number of ways to prove this. We give below a simple proof by induction. Let's begin 
by writing explicitly the expressions for $A_1, A_2$:
\begin{subequations} \begin{align}
A_1 & = (1-P_1)\\
A_2 & = P_1(1-P_2)= P_1 - P_1P_2
\end{align} \end{subequations}
Note that $P_1 = 1-A_1$, and so we may rewrite $A_2$ as
\begin{equation}
A_2 = 1-A_1 - P_1P_2
\end{equation}
from which we derive that
\begin{equation} \label{eq:A_1+A_2}
A_1 + A_2 = 1 - P_1P_2 \ .
\end{equation}
If we now consider $A_3$, we have
\begin{equation} \label{eq:A_3}
A_3 = P_1P_2(1-P_3) = P_1P_2 - P_1P_2P_3 \ .
\end{equation}
Since, from Eq. \eqref{eq:A_1+A_2}, $P_1P_2 = 1-A_1-A_2$, then we may also rewrite 
Eq. \eqref{eq:A_3} as,
\begin{equation}
A_3 = 1-A_1-A_2 - P_1P_2P_3 \ ,
\end{equation}
from which we also derive that
\begin{equation}
A_1+A_2+A_3 =  1 - P_1P_2P_3 \ .
\end{equation}
Following the same line of reasoning, we can see that, in general,
\begin{equation}
\sum^n_{m=1} A_m = 1 - \prod^n_{m=1} P_i \ .
\end{equation}

\section{Initial molecular structures (Cartesian coordinates in Angstroms)}

\begin{verbatim}

N2 R_N-N = 1.067 Angstroms:
    I     Atom           X                Y                Z
 ----------------------------------------------------------------
    1      N       0.0000000000     0.0000000000     0.0000000000
    2      N       0.0000000000     0.0000000000     1.0671420000
 ----------------------------------------------------------------
 
N2 R_N-N = 1.400 Angstroms:
    I     Atom           X                Y                Z
 ----------------------------------------------------------------
    1      N       0.0000000000     0.0000000000     0.0000000000
    2      N       0.0000000000     0.0000000000     1.4000000000
 ----------------------------------------------------------------
 
N2 R_N-N = 1.150 Angstroms: 
    I     Atom           X                Y                Z
 ----------------------------------------------------------------
    1      N       0.0000000000     0.0000000000     0.0000000000
    2      N       0.0000000000     0.0000000000     1.1500000000
 ----------------------------------------------------------------

Ethylene:
    I     Atom           X                Y                Z
 ----------------------------------------------------------------
    1      C       0.0000000000     0.0000000000     0.0000000000
    2      H       0.0000000000     0.0000000000     1.0742160000
    3      H       0.9595703355     0.0000000000    -0.4828713968
    4      C      -1.1191466626     0.0000000000    -0.6896850753
    5      H      -2.0787169982     0.0000000000    -0.2068136786
    6      H      -1.1191466626     0.0000000000    -1.7639010753
 ----------------------------------------------------------------
 
 
 
 
Chloroethylene:
     I     Atom           X                Y                Z
 ----------------------------------------------------------------
    1      Cl      0.0000000000     0.0000000000     0.0000000000
    2      C       0.0000000000     0.0000000000     1.7316720000
    3      H       0.9853392606     0.0000000000     2.1483176034
    4      C      -1.0935317241    -0.0000000000     2.4493744736
    5      H      -2.0698140500    -0.0000000000     2.0064002052
    6      H      -1.0220874542    -0.0000000000     3.5199672631
 ----------------------------------------------------------------
 
cis-dichloroethylene:
     I     Atom           X                Y                Z
 ----------------------------------------------------------------
    1      Cl      0.0000000000     0.0000000000     0.0000000000
    2      C       0.0000000000     0.0000000000     1.7158570000
    3      H       0.9754145626     0.0000000000     2.1546605495
    4      C      -1.0642674138    -0.0000000000     2.4775763725
    5      H      -0.9635098987    -0.0000000000     3.5423909325
    6      Cl     -2.6884303018    -0.0000000000     1.9241686045
 ----------------------------------------------------------------
 
trans-dichloroethylene:
     I     Atom           X                Y                Z
 ----------------------------------------------------------------
    1      Cl      0.0000000000     0.0000000000     0.0000000000
    2      C       0.0000000000     0.0000000000     1.7251020000
    3      H       0.9725182492     0.0000000000     2.1682755046
    4      C      -1.1108641214    -0.0000000000     2.4124445514
    5      H      -2.0833803441    -0.0000000000     1.9692665997
    6      Cl     -1.1108732143     0.0000000000     4.1375455514
 ----------------------------------------------------------------
 
1,1-dichloroethylene:
     I     Atom           X                Y                Z
 ----------------------------------------------------------------
    1      Cl      0.0000000000     0.0000000000     0.0000000000
    2      C       0.0000000000     0.0000000000     1.7223030000
    3      C       1.0985662562     0.0000000000     2.4310014204
    4      H       1.0546069317     0.0008291316     3.5008653643
    5      H       2.0551413131    -0.0005294646     1.9498419721
    6      Cl     -1.5693772271     0.0020344528     2.4319816909
 ----------------------------------------------------------------
 
 
trichloroethylene:
     I     Atom           X                Y                Z
 ----------------------------------------------------------------
    1      Cl      0.0000000000     0.0000000000     0.0000000000
    2      C       0.0000000000     0.0000000000     1.7092420000
    3      C       1.0751798641     0.0000000000     2.4583761272
    4      H       0.9959360068     0.0004721400     3.5236907956
    5      Cl      2.6789106202    -0.0002059649     1.8573907787
    6      Cl     -1.5586905277     0.0010844173     2.4407910370
 ----------------------------------------------------------------
 
tetrachloroethylene:
     I     Atom           X                Y                Z
 ----------------------------------------------------------------
    1      Cl      0.0000000000     0.0000000000     0.0000000000
    2      C       0.0000000000     0.0000000000     1.7227670000
    3      C       1.1127702200     0.0000000000     2.4402796195
    4      Cl      2.6820100527    -0.0002426958     1.7293487076
    5      Cl      1.1129142371     0.0003344598     4.1630555810
    6      Cl     -1.5693071217     0.0005641705     2.4335709971
 ----------------------------------------------------------------
 
chloroethane:
     I     Atom           X                Y                Z
 ----------------------------------------------------------------
    1      Cl      0.0000000000     0.0000000000     0.0000000000
    2      C       0.0000000000     0.0000000000     1.7959090000
    3      H       1.0352683828     0.0000000000     2.0949364710
    4      H      -0.4531548755     0.9308229065     2.0949364710
    5      C      -0.7464826193    -1.1936557262     2.3509867800
    6      H      -0.2904814026    -2.1205141302     2.0296230487
    7      H      -1.7794323605    -1.1893612135     2.0296230487
    8      H      -0.7229945558    -1.1560973682     3.4350270623
 ----------------------------------------------------------------
 
 \end{verbatim}
 

\section{CAP box parameters}
As mentioned in the article, the CAP Hamiltonian, $H^\eta$, is given by \cite{riss93,sommerfeld98,
santra01,zuev14,dempwolff22}
\begin{equation} \label{eq:H_eta}
H^\eta = H - i \eta W \ ,
\end{equation}
where $H$ is the physical Hamiltonian, $\eta$ is a real scalar parameter and $W$ is a Hermitian 
operator. The potential $W$ was chosen to be quadratic of the following form,
\begin{equation} \label{eq:W_alpha_quadratic}
W = \sum_{\alpha = x,y,z} \ W_\alpha \ , \qquad W_\alpha = \theta(\Delta_\alpha - r^o_\alpha) \cdot 
(\Delta_\alpha - r^o_\alpha)^2 \ , \ \Delta_\alpha \equiv \abs{r_\alpha - o_\alpha}
\end{equation}
where $\theta(x)$ is a piecewise function defined as
\begin{equation}
\theta(x) \equiv \begin{cases} 1 \ & \text{if } x >0 \\ 0 \ & \text{if } x \leq 0
\end{cases} \ .
\end{equation}
$r_\alpha$ is the electronic coordinate along axis $\alpha$. The vector $(o_x,o_y,o_z)$ is the origin 
of the CAP, which in our calculations is chosen to coincide with the center of nuclear charge \cite{benda17}, 
that is 
\begin{equation}
o_\alpha = \frac{\sum_k R_{k,\alpha} Z_k}{\sum_k Z_k} \ ,
\end{equation}
where $Z_k$ and $R_k$ are the charge and coordinate of the $k-$th nucleus, respectively. This choice 
of $o_\alpha$ allows the origin of the CAP to move with the nuclear frame during the AIMD simulation. 

Similar to single point energy CAP calculations, to run a CAP-AIMD simulation, one needs to specify the 
values of the set $\{r^o_\alpha, \eta\}$, which parameterize the CAP. It has been advocated \cite{jagau14,
zuev14,benda17} to set $r^o_\alpha $ for temporary radical anions equal to the second moment of the 
neutral molecule's wave function along the axis $\alpha$, i.e. $r^o_\alpha = \sqrt{\left<{r_\alpha^2}\right>}$. 
To minimize the perturbation of the resonance wave function by the CAP, one runs an $\eta-$trajectory 
to determine an optimal $\eta$, i.e. $\eta_{opt}$, which is chosen such that \cite{riss93}
\begin{equation}
\eta_{opt} = \min_\eta \abs{\eta \frac{dE}{d\eta}}  \ .
\end{equation}

However, in our calculations, we resorted to a more complete approach, whereby the CAP box 
dimensions, $r^o_\alpha$, are also optimized. The details of the approach will be expounded on in 
an upcoming work. For now, it suffices to say that the optimized CAP parameters, $\{r^o_{\alpha,opt}, 
\eta_{opt}\}$, were chosen according to criterion
\begin{equation}
\{r^o_{\alpha,opt}, \eta_{opt}\} = \min_{\{r^o_{\alpha}, \, \eta_{opt}\}} \abs\Big{\Im \left<-i \eta W \right>} \ .
\end{equation}

The optimized CAP parameters used in our CAP-AIMD simulations are summarized in Tab. \ref{tbl:CAP}. 
The vertical attachment energies (VAE) computed using these parameters are reported in Tab. 
\ref{tbl:energy} and compared with results from static CAP-EOM-EA-CCSD calculations taken from 
Ref. \citenum{benda18b}. It is interesting to note that the ratio between the CAP-EOM-EA-CCSD 
VAE's and those from CAP-HF is always between 0.6 and 0.7.

In Tab. \ref{tbl:energy}, we also report values for $\abs{\Re \left<-i \eta W \right> / \, \text{VAE}}$ and 
$\abs{2 \Im \left<-i \eta W \right> / \, \Gamma}$ from our CAP-HF calculations. These ratios may be 
interpreted as relative errors in the computed attachment energies and resonance widths, respectively. 
The closer they are to zero, the better. 

\begin{table} 
\caption{CAP parameters optimized at the equilibrium geometry of the neutral molecule, used in 
the CAP-AIMD simulations. Basis set is cc-pVTZ+3p for all molecules except \ch{C2Cl4} where 
aug-cc-pVDZ+3p is used. The extra diffuse functions are placed on all atoms except H.}
\begin{tabular}{lllll} \hline
Molecule & $\eta_{opt}$\textsuperscript{\emph{a}} & $r^o_{x,opt}$\textsuperscript{\emph{b}} & 
$r^o_{y,opt}$  & $r^o_{z,opt}$ \\ \hline
\ch{N2} & 2490 & 3535 & 3535 & 8102 \\
\ch{C2H5Cl} &  650 & 3994 & 4148 & 4163 \\ 
\ch{C2H4} & 230 & 4360 & 2680  & 4360 \\
\ch{C2H3Cl} & 460 & 8097 & 4490 & 16823 \\
\textit{cis}-\ch{C2H2Cl2} & 500  & 8300 & 6500 & 8300 \\
\textit{trans}-\ch{C2H2Cl2} & 500 & 6388 & 5355 & 32786 \\
1,1-\ch{C2H2Cl2} & 10 & 23800 & 10856 & 25131 \\
\ch{C2HCl3} & 500 & 7880 & 6200 & 7880 \\
\ch{C2Cl4} & 2340 & 28080 & 7840 & 43774 \\ \hline
\end{tabular} 
\begin{flushleft}
\textsuperscript{\emph{a}} $\eta$ values are in $\times 10 ^{-5} $ a.u.\\
\textsuperscript{\emph{b}} $r^o$ values are in $\times 10^{-3}$ a.u. (bohr). 
\end{flushleft} \label{tbl:CAP}
\end{table}

\begin{table}
\caption{Vertical Attachment Energies (VAE) and resonance widths ($\Gamma$) in eV of 
dinitrogen, chloroethane, ethylene and the mono-, di-, tri- and tetra- chloro-substituted ethylenes 
computed at the optimized geometry of the neutral molecules with CAP-HF. Values computed 
with CAP-EOM-EA-CCSD and experimental values are also shown.}
\begin{tabular}{llllllll} \hline 
Molecule & expt \textsuperscript{\emph{a}} & \multicolumn{2}{c}{CAP-EOM-EA-CCSD \textsuperscript{\emph{b}}} & 
\multicolumn{4}{c}{CAP-HF} \\
 & VAE & VAE & $\Gamma$ & VAE  & $\Gamma$ & \small{$\abs{\frac{\Re \left<-i \eta W \right>}{\text{VAE}}} $} & 
 \small{$\abs{\frac{2\Im \left<-i \eta W \right>}{\Gamma}}$} \\ \hline
\ch{N2} & 2.3 & -- & -- & 3.396 & 0.362 &  $1.91\cdot 10^{-2}$  & $4.08 \cdot 10^{-4}$ \\
\ch{C2H5Cl} & 2.35 & -- & -- & 4.033 & 1.488 & $2.67\cdot 10^{-3}$ & $1.83 \cdot 10^{-3}$ \\ 
\ch{C2H4} & 1.73 & 2.155 & 0.446 & 2.908 & 0.661 & $2.30 \cdot 10^{-2}$ & $1.27\cdot 10^{-3}$ \\
\ch{C2H3Cl} & 1.28 & 1.730 & 0.266	& 2.368 & 0.341 & $4.91 \cdot 10^{-3}$ & $1.11 \cdot 10^{-2}$ \\
\textit{cis}-\ch{C2H2Cl2} & 1.11 & 1.573 & 0.056 & 2.158 & 0.015 & $4.41 \cdot 10^{-4}$  & $6.32 \cdot 10^{-5}$ \\
\textit{trans}-\ch{C2H2Cl2} & 0.80 & 1.335 & 0.109 & 1.942 & 0.059 & $1.68\cdot 10^{-3}$  & $1.26 \cdot 10^{-3}$ \\
1,1-\ch{C2H2Cl2} & 0.76 & 1.285 & 0.245	& 1.858 & 0.535 & $5.37 \cdot 10^{-6}$ & $4.29\cdot 10^{-4}$ \\
\ch{C2HCl3} & 0.59 & 1.052 & 0.169 & 1.637 & 0.034 & $1.48 \cdot 10^{-3}$ & $1.12 \cdot 10^{-3}$ \\
\ch{C2Cl4} & 0.3 & 0.982 & 0.026 & 1.286 & 0.007 & $8.51\cdot 10^{-4}$ & $8.23\cdot 10^{-2}$ \\ \hline
\end{tabular} 
\begin{flushleft}
\textsuperscript{\emph{a}} From Ref. \citenum{burrow81}. \\
\textsuperscript{\emph{b}} From Ref. \citenum{benda18b}.
\end{flushleft} \label{tbl:energy}
\end{table}

\newpage


\section{Results}
Below, we show scatter plots related to different specific trajectories in different grades of gray. 
Average i) real part of the complex SCF energy, ii) resonance width, and iii) lost fraction at each 
time step are shown in i) dark orange, ii) red,  and iii) blue dashed lines, respectively. To make 
some graphs more intelligible, we plot the related data only for some few initial steps. Unless 
otherwise stated, the initial geometry is always the optimized geometry of the neutral molecule.
All trajectories were run in the microcanonical ensemble; the reported temperatures define the 
Maxwell-Boltzmann distribution from which the initial velocities were sampled.

\subsection{Dinitrogen, \ch{N2}}
The dinitrogen anion was simulated at three different initial geometries: 
i) $R^o_{N-N}=1.067$ \AA, ii) $R^o_{N-N}=1.150$ \AA\ and iii) $R^o_{N-N}=1.400$ \AA. 
The time step chosen for the simulation does not matter as long as the condition 
$ \Gamma_0 \Delta t < 1 $ is respected, as explained in the article.

\subsubsection{Dinitrogen, \ch{N2}, $R^o_{N-N}=1.067$ \AA}
Time step = 10 a.u. (0.2419 fs); 
Number of steps = 1500; 
Temperature = 300 K;
Number of trajectories = 200;
$\eta = 0.02490$ a.u.

\begin{figure}[H]
\includegraphics[scale=0.8]{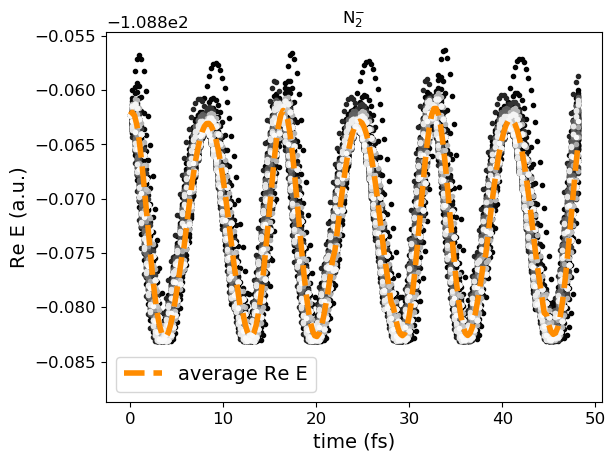} 
\caption{$\Re E$ v time for dinitrogen anion. $R^o_{N-N}=1.067$ \AA. An offset of --108.8 is used for the y-axis.}
\label{fgr:1_ReE}
\end{figure}

\begin{figure}[H]
\includegraphics[scale=0.8]{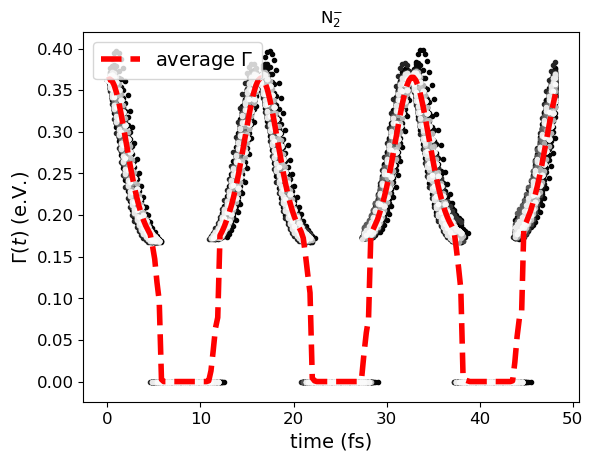} 
\caption{Resonance width, $\Gamma$, v time for dinitrogen anion. $R^o_{N-N}=1.067$ \AA}
\label{fgr:1_Gamma}
\end{figure}

\begin{figure}[H]
\includegraphics[scale=0.8]{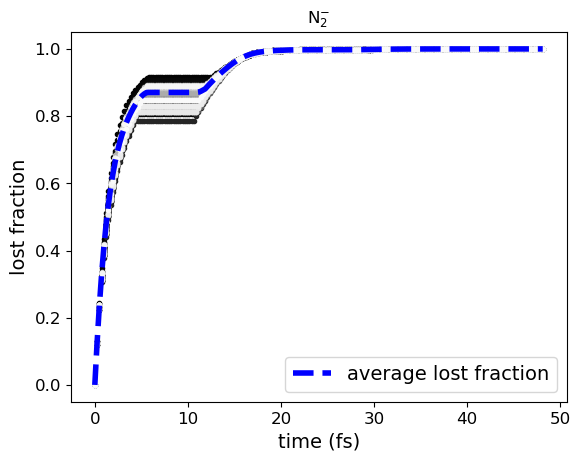} 
\caption{Lost fraction v time for dinitrogen anion. $R^o_{N-N}=1.067$\AA. The equilibrium 
lost fraction is 1.0 for all trajectories.}
\label{fgr:1_lost_fraction}
\end{figure}

\subsubsection{Dinitrogen, \ch{N2}, $R^o_{N-N}=1.400$ \AA}
Time step = 2 a.u. (0.0484 fs); 
Number of steps = 2000; 
Temperature = 300 K;
Number of trajectories = 40;
$\eta = 0.00500$ a.u.

\begin{figure}[H]
\includegraphics[scale=0.8]{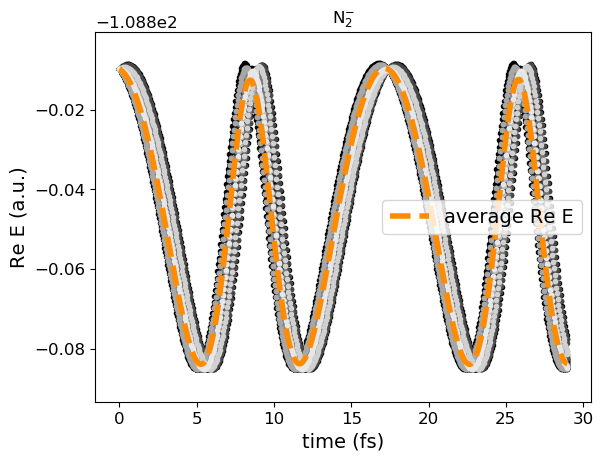} 
\caption{$\Re E$ v time for dinitrogen anion. $R^o_{N-N}=1.400$\AA. An offset of --108.8 is used for the y-axis.}
\label{fgr:1_140pm_ReE}
\end{figure}

\begin{figure}[H]
\includegraphics[scale=0.8]{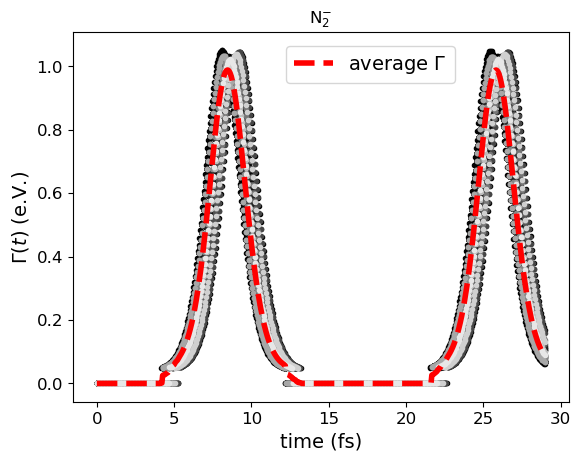} 
\caption{Resonance width, $\Gamma$, v time for dinitrogen anion. $R^o_{N-N}=1.400$ \AA }
\label{fgr:1_140pm_Gamma}
\end{figure}

\begin{figure}[H]
\includegraphics[scale=0.8]{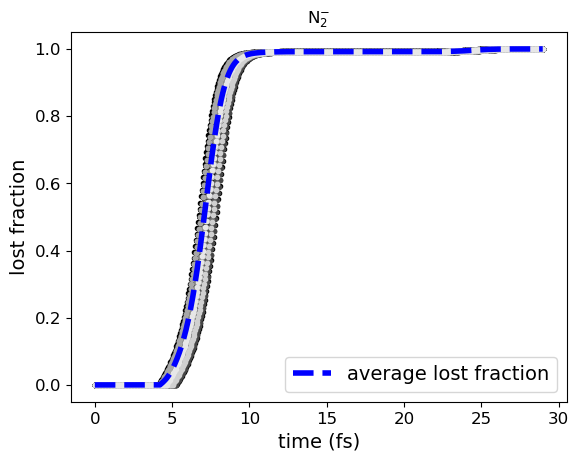} 
\caption{Lost fraction v time for dinitrogen anion. $R^o_{N-N}=1.400$ \AA. The equilibrium 
lost fraction is 1.0 for all trajectories.}
\label{fgr:1_140pm_lost_fraction}
\end{figure}

\subsubsection{Dinitrogen, \ch{N2}, $R^o_{N-N}=1.150$ \AA.}
Time step = 2 a.u. (0.0484 fs); 
Number of steps = 2000; 
Temperature = 300 K;
Number of trajectories = 51;
$\eta = 0.00070$ a.u.

\begin{figure}[H]
\includegraphics[scale=0.8]{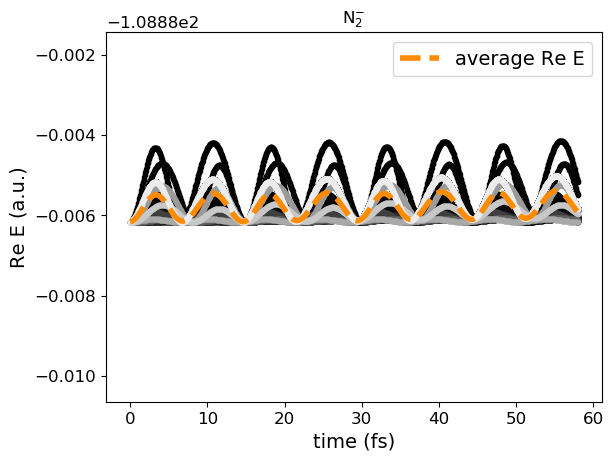} 
\caption{$\Re E$ v time for dinitrogen anion. $R^o_{N-N}=1.150$ \AA. An offset of --108.8 is used for the y-axis.}
\label{fgr:1_115pm_ReE}
\end{figure}

\begin{figure}[H]
\includegraphics[scale=0.8]{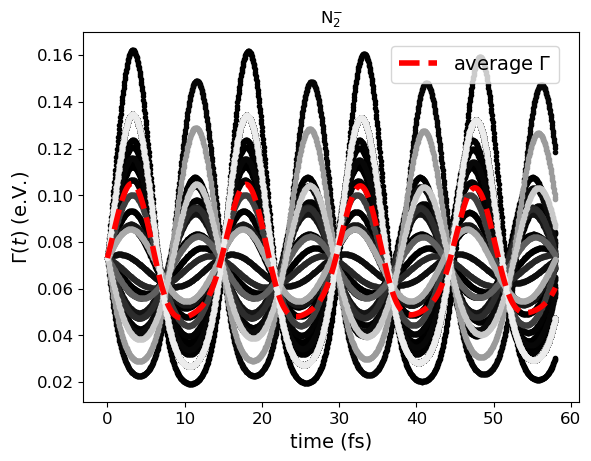} 
\caption{Resonance width, $\Gamma$, v time for dinitrogen anion. $R^o_{N-N}=1.150$ \AA.}
\label{fgr:1_115pm_Gamma}
\end{figure}

We can see in these simulations that the $\Gamma$ time profile (Fig. \ref{fgr:1_115pm_Gamma}) 
is not uniform as in the other initial geometries of the same anion considered above; in particular, 
it could increase or decrease in the beginning. This is related to the high dependence of $\Gamma$ 
on the initial velocities at this initial geometry, which is in the neighborhood of the anion's equilibrium 
structure. The plot of the real part of the complex SCF energy, Re$ \ E$, against time (Fig. 
\ref{fgr:1_115pm_ReE}) also shows a small difference between the maxima and minima of the peaks 
compared to the other initial geometries considered above. Just like the $\Gamma$ profile (Fig. 
\ref{fgr:1_115pm_Gamma}) the lost fraction time profile also shows less homogeneity compared 
to Figs. \ref{fgr:1_lost_fraction} and \ref{fgr:1_140pm_lost_fraction}.

\begin{figure}[H]
\includegraphics[scale=0.8]{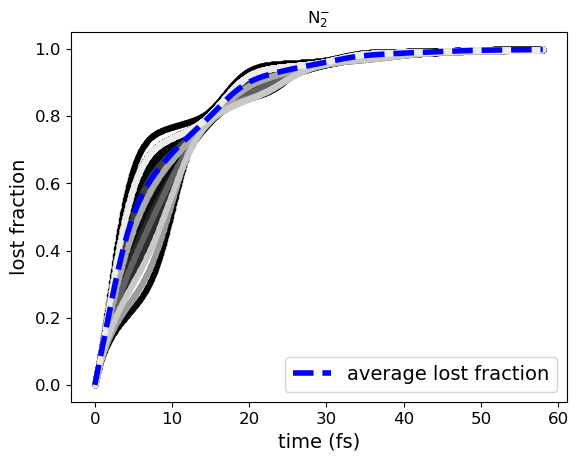} 
\caption{Lost fraction v time for dinitrogen anion. $R^o_{N-N}=1.150$ \AA. The equilibrium 
lost fraction is 1.0 for all trajectories.}
\label{fgr:1_115pm_lost_fraction}
\end{figure}

\subsubsection{Dinitrogen, variation of $\Gamma$ with N-N bond length}
\begin{figure}[H]
\includegraphics[scale=0.5]{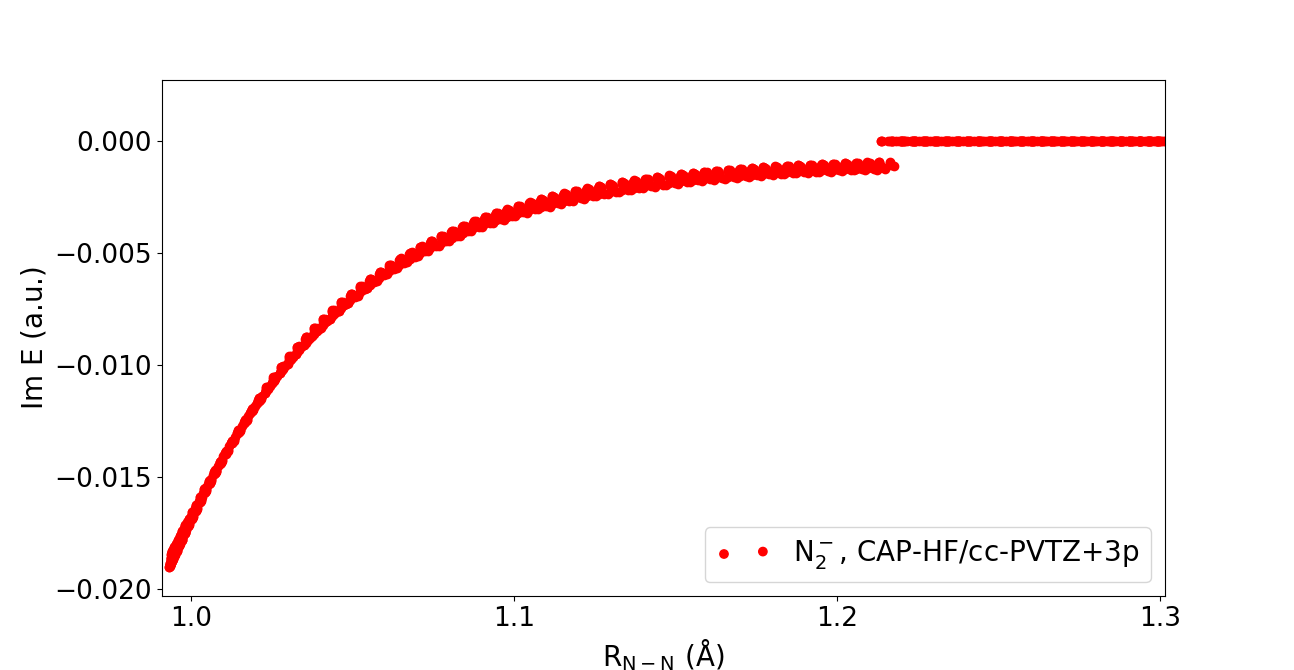} 
\caption{The plot shows variation of the imaginary part of the CPES of \ch{N2-} with N-N bond length, $R_{N-N}$. The points derive from a CAP-AIMD simulation performed at a time step 
of 2 a.u. $\approx$ 0.05 fs, for a total of 2000 time steps.}
\label{fgr:N2PES_Im}
\end{figure}

In Fig. 2 of the article, we reported the real part of the CPES of \ch{N2-} obtained from CAP-AIMD simulation. We now report in Fig. \ref{fgr:N2PES_Im} the imaginary part of the same CPES.

\par Starting from the compressed N-N bond length of $R_{N-N} \sim 1.0$\AA, we see that $\Im E$ increases as $R_{N-N}$ increases. The upper limit of $\Im E$ is zero, which is associated to bound \ch{N2-}. 

\par We observe in Fig. \ref{fgr:N2PES_Im} a discontinuity in $\Im E$ as $\Im E \to 0$, which is around $R_{N-N} \approx 1.2$\AA. The discontinuity is an indication of the transition between bound and unbound \ch{N2-}. It is also a consequence of the limitations of the CAP-HF method. 

\subsection{Ethylene, \ch{C2H4}}
Time step = 5 a.u. (0.1209 fs); 
Number of steps = 1000; 
Temperature = 200 K;
Number of trajectories = 200;
$\eta = 0.00230$ a.u.

Looking at the actual trajectories (in shades of gray) in Fig. \ref{fgr:3_Gamma}, we see that 
$\Gamma$ becomes zero intermittently, indicating the extra electron has become bound; the 
average $\Gamma$ (in red) on the other hand, never becomes zero. This is due to the large 
variance among the trajectories in regards to the $\Gamma$ profile. However, as we can see 
from Fig. \ref{fgr:3_lost_fraction}, there is a great homogeneity among the trajectories in regards 
to the lost fraction's time profile. This is due to the high homogeneity we see in the $\Gamma$ 
profiles in the first 10fs.

\begin{figure}[h!]
\includegraphics[scale=0.8]{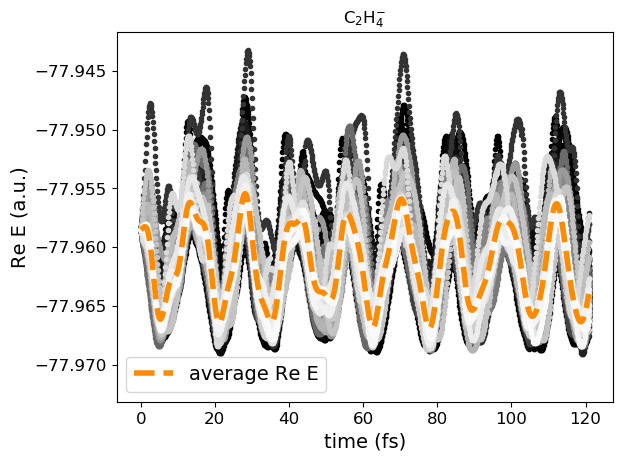} 
\caption{$\Re E$ v time for ethylene anion.}
\label{fgr:3_ReE}
\end{figure}

\begin{figure}[h!]
\includegraphics[scale=0.8]{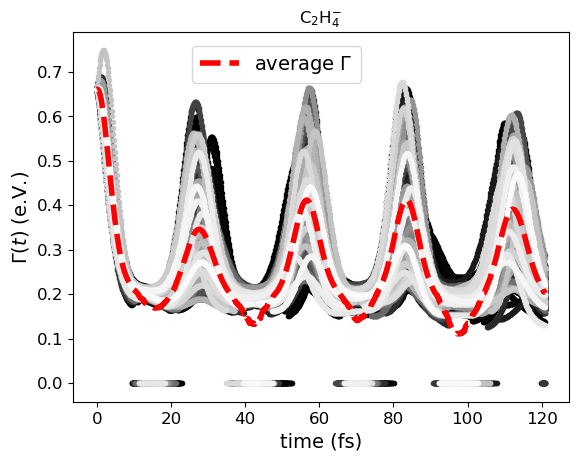} 
\caption{Resonance width, $\Gamma$, v time for ethylene anion.}
\label{fgr:3_Gamma}
\end{figure}

\clearpage

\begin{figure}
\includegraphics[scale=0.8]{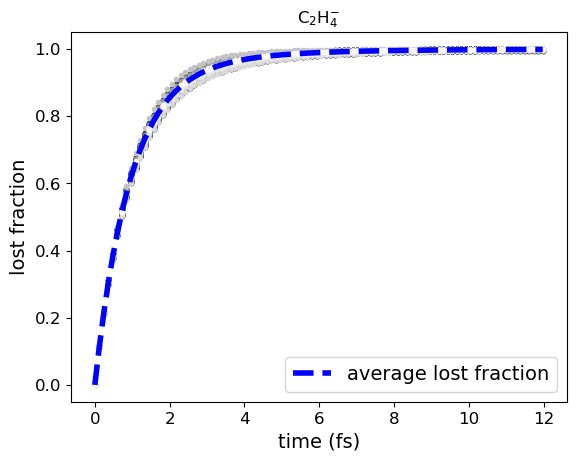} 
\caption{Lost fraction v time for ethylene anion. The equilibrium lost fraction is $1.0$ across 
all trajectories.}
\label{fgr:3_lost_fraction}
\end{figure}


\subsection{Chloroethylene, \ch{C2H3Cl}}
Time step = 10 a.u. (0.2419 fs); 
Number of steps = 1000; 
Temperature = 300 K;
Number of trajectories = 200;
$\eta = 0.00460$ a.u.

Statistics on equilibrium lost fraction: \textit{minimum}= $0.7480$; \textit{maximum}=$0.9973$; 
\textit{mean} =$0.9220$; \textit{median}=$0.9305$; \textit{standard deviation} = $0.0416$.

In all trajectories, we observe an initial stage where $\Gamma$ decreases as the nuclei relax. 
In some trajectories, that stage is followed by a stage where $\Gamma$ increases again due to 
the nuclear configurations reached, before eventually becoming zero; this gives rise to the curve 
centered around 25 fs in Fig. \ref{fgr:5_Gamma}.

\begin{figure}[H]
\includegraphics[scale=0.8]{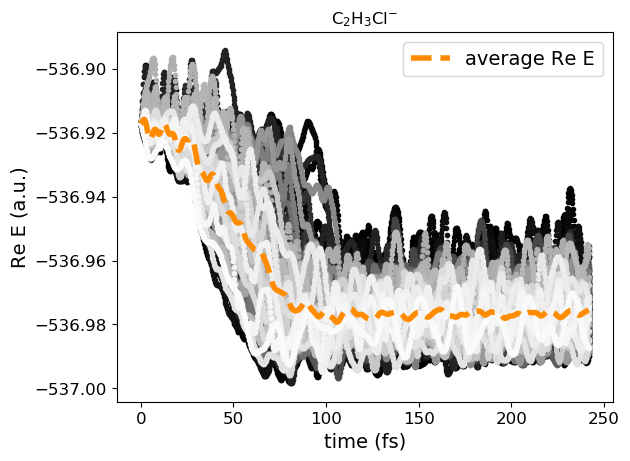} 
\caption{$\Re E$ v time for chloroethylene anion.}
\label{fgr:5_ReE}
\end{figure}

\begin{figure}[H]
\includegraphics[scale=0.8]{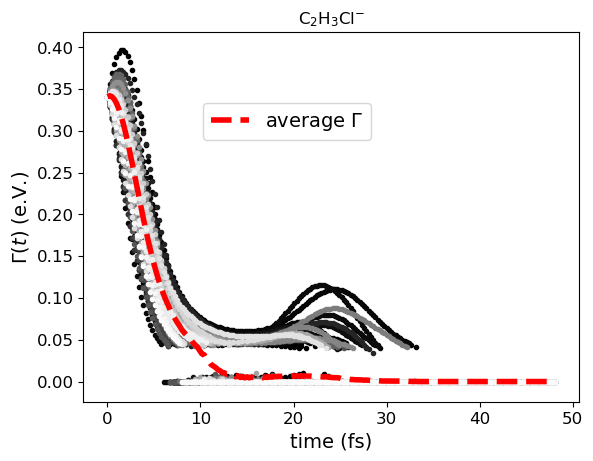} 
\caption{Resonance width, $\Gamma$, v time for chloroethylene anion.}
\label{fgr:5_Gamma}
\end{figure}

\begin{figure}[H]
\includegraphics[scale=0.8]{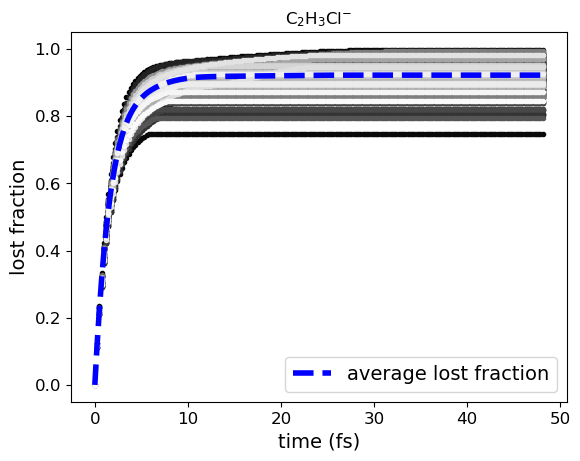} 
\caption{Lost fraction v time for chloroethylene anion.}
  \label{fgr:5_lost_fraction}
\end{figure}

\subsubsection{What happens when the wrong CAP parameters are used?}
To illustrate that it is important to run CAP-AIMD simulations with the right CAP parameters, 
we show below results for chloroethylene anion using the same initial geometry and CAP box 
parameters used for the simulations above (and listed in Tab. \ref{tbl:CAP}), but this time with 
$\eta = 0.010$ a.u. We chose this value for $\eta$ because the $\eta-$trajectory (with the 
same box dimensions as in Tab. \ref{tbl:CAP}) indicates that the electronic state of the anion 
at this $\eta$ value has pseudocontinuum character and is not a resonance.

\begin{figure}[H]
\includegraphics[scale=0.8]{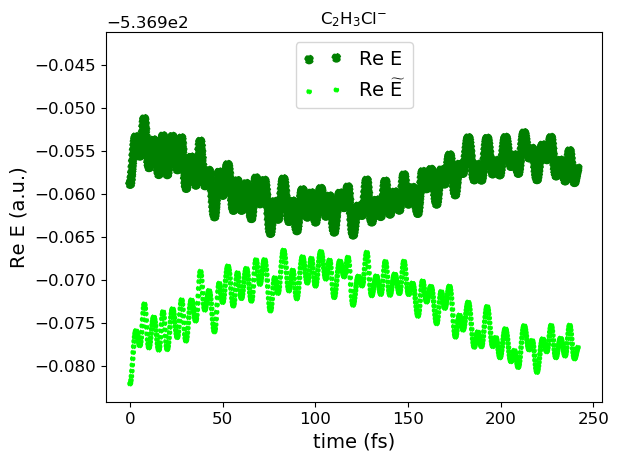} 
\caption{$\Re E$ and $\Re \widetilde{E}$ v time for chloroethylene anion using $\eta=0.010$ a.u. 
An offset of --536.9 is used for the y-axis.}
\label{fgr:5_continuum_ReE}
\end{figure}

\begin{figure}[H]
\includegraphics[scale=0.8]{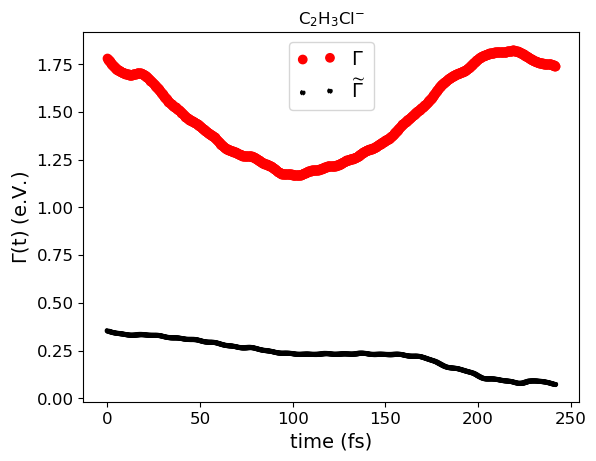} 
\caption{$\Gamma$ and $\widetilde{\Gamma}$ v time for chloroethylene anion using $\eta=0.010$ a.u.}
\label{fgr:5_continuum_Gamma}
\end{figure}

Fig. \ref{fgr:5_continuum_ReE} and \ref{fgr:5_continuum_Gamma} show disagreement 
between $\Re E$, $\Gamma$ and their respective deperturbed counterparts. In Fig. 
\ref{fgr:chloroethylenes_etc_and_ReE_overlap}, we show as a counterexample, the 
$\Re E$ and $\Re \widetilde{E}$ values for the trajectories discussed in Fig. 3 of the article; 
in Fig. \ref{fgr:1_ReE_overlap} we also show the $\Re E$ and $\Re \widetilde{E}$ values 
for the trajectories discussed in Fig. 1 of the article for \ch{N2-}. The neat agreement 
between $\Re E$, $\Gamma$ and their respective deperturbed counterparts in Fig. 
\ref{fgr:chloroethylenes_etc_and_ReE_overlap} and \ref{fgr:1_ReE_overlap} is in 
sharp contrast to what we see in Fig. \ref{fgr:5_continuum_ReE} and \ref{fgr:5_continuum_Gamma}.  

\begin{figure}[H]
\includegraphics[scale=0.8]{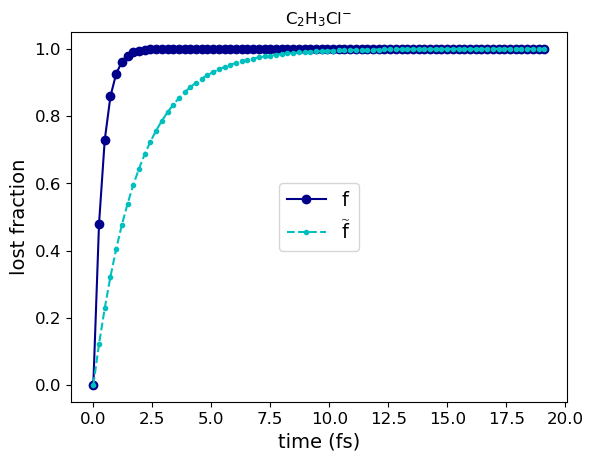} 
\caption{Lost fraction $f$ and deperturbed $\widetilde{f}$ v time for chloroethylene anion 
using $\eta=0.010$ a.u.}
\label{fgr:5_continuum_lost_fraction}
\end{figure}

\begin{figure}[H]
\includegraphics[scale=0.5]{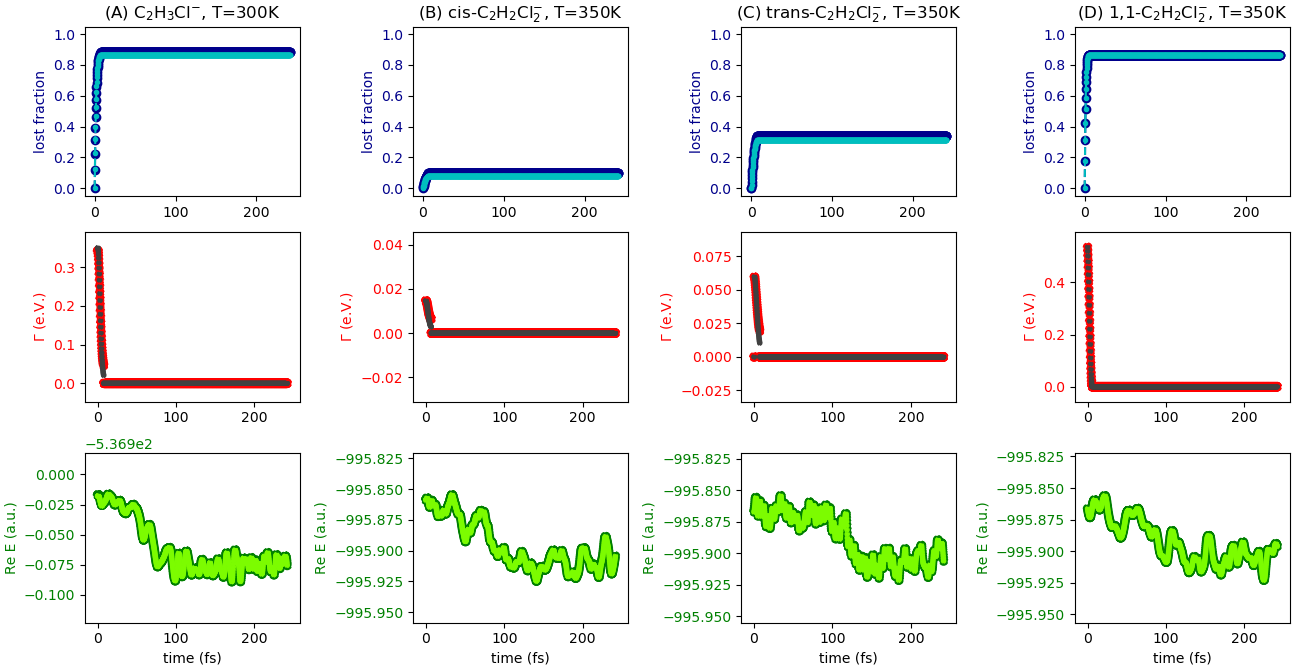} 
\caption{Lost fraction, $f$, $\Re E$ and resonance width, $\Gamma$, profiles and their 
deperturbed counterparts, $\widetilde{f}$, $\Re \widetilde{E}$ and $\widetilde{\Gamma}$ for 
randomly chosen CAP-AIMD trajectories of A) chloroethylene, B) \textit{cis-}, C) \textit{trans-} 
and D) \textit{1,1}-dichloroethylene anions, simulated with the appropriate CAP parameters 
from Tab. \ref{tbl:CAP}.}
\label{fgr:chloroethylenes_etc_and_ReE_overlap}
\end{figure}

\begin{figure}[H]
\includegraphics[scale=0.5]{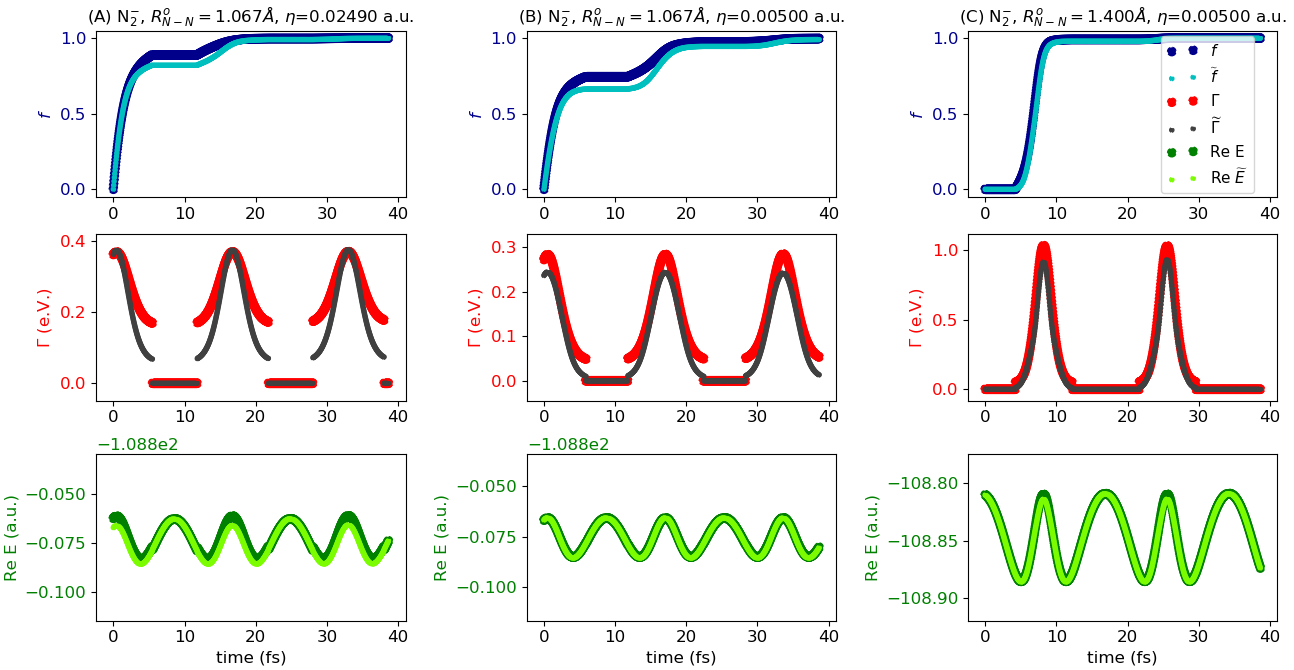} 
\caption{Lost fraction, $f$, $\Re E$ and resonance width, $\Gamma$, profiles (and their 
deperturbed counterparts, $\widetilde{f}$, $\Re \widetilde{E}$ and $\widetilde{\Gamma}$) 
for randomly chosen CAP-AIMD trajectories of \ch{N2-}, simulated with the appropriate CAP 
parameters from Tab. \ref{tbl:CAP}.}
\label{fgr:1_ReE_overlap}
\end{figure}


\subsection{\textit{cis}-dichloroethylene, \ch{C2H2Cl2}}
Time step = 10 a.u. (0.2419 fs); 
Number of steps = 1000; 
Temperature = 350 K;
Number of trajectories = 199;
$\eta = 0.00500$ a.u.

Statistics on equilibrium lost fraction: \textit{minimum}= $0.0606$; \textit{maximum}=$0.1736$; 
\textit{mean} =$0.0995$; \textit{median}=$0.0981$; \textit{standard deviation} = $0.0168$.

\begin{figure}[H]
\includegraphics[scale=0.8]{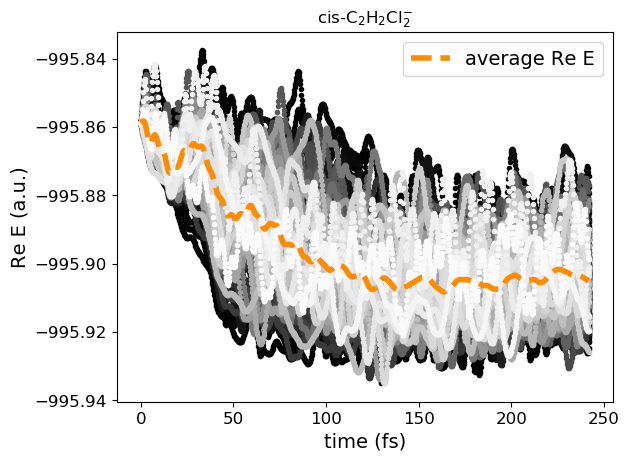} 
\caption{$\Re E$ v time for \textit{cis}-dichloroethylene anion.}
\label{fgr:6_ReE}
\end{figure}

\begin{figure}[H]
\includegraphics[scale=0.8]{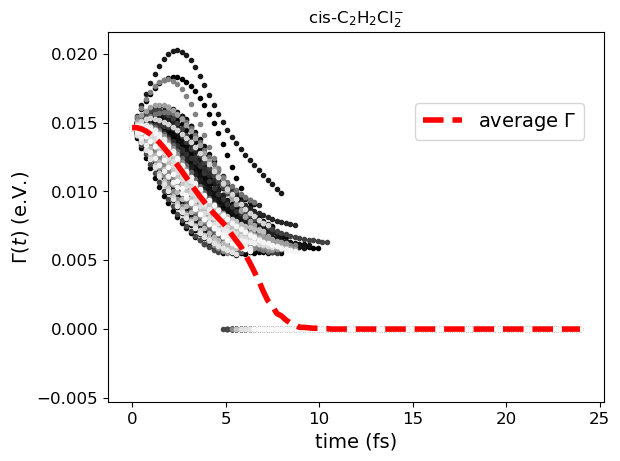} 
\caption{Resonance width, $\Gamma$, v time for \textit{cis}-dichloroethylene anion. }
\label{fgr:6_Gamma}
\end{figure}

\begin{figure}[H]
\includegraphics[scale=0.8]{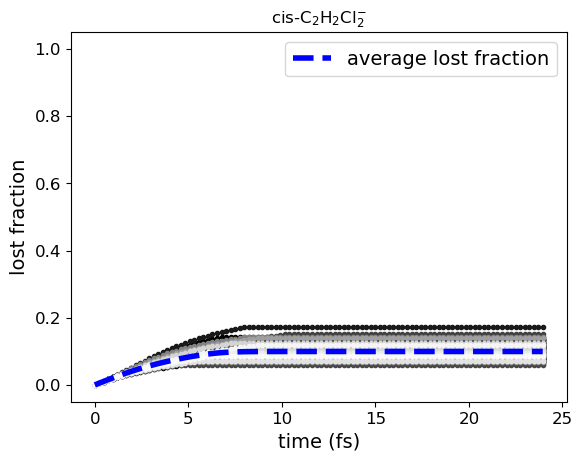} 
\caption{Lost fraction v time for \textit{cis}-dichloroethylene anion.}
\label{fgr:6_lost_fraction}
\end{figure}


\subsection{\textit{trans}-dichloroethylene, \ch{C2H2Cl2}}
Time step = 10 a.u. (0.2419 fs); 
Number of steps = 1000; 
Temperature = 350 K;
Number of trajectories = 200;
$\eta = 0.00500$ a.u.

Statistics on equilibrium lost fraction: \textit{minimum}= $0.2089$; \textit{maximum}=$0.4473$; 
\textit{mean} =$0.3264$; \textit{median}=$0.3294$; \textit{standard deviation} = $0.0503$.

\begin{figure}[H]
\includegraphics[scale=0.8]{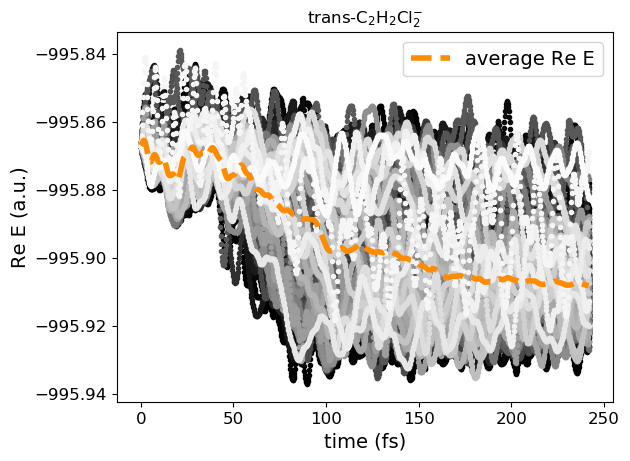} 
\caption{$\Re E$ v time for \textit{trans}-dichloroethylene anion.}
\label{fgr:7_ReE}
\end{figure}

\begin{figure}[H]
\includegraphics[scale=0.8]{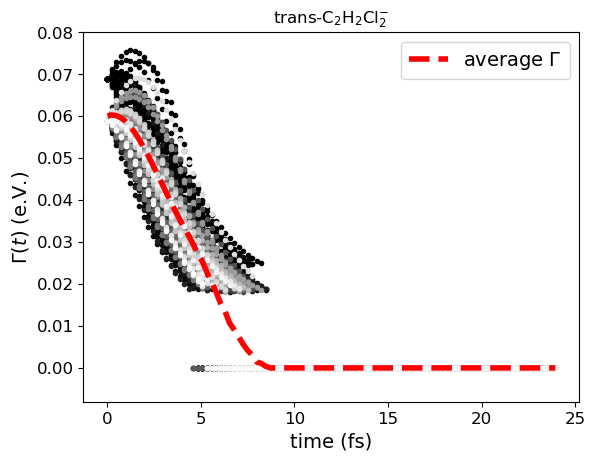} 
\caption{Resonance width, $\Gamma$, v time for \textit{trans}-dichloroethylene anion.}
\label{fgr:7_Gamma}
\end{figure}

\begin{figure}[H]
\includegraphics[scale=0.8]{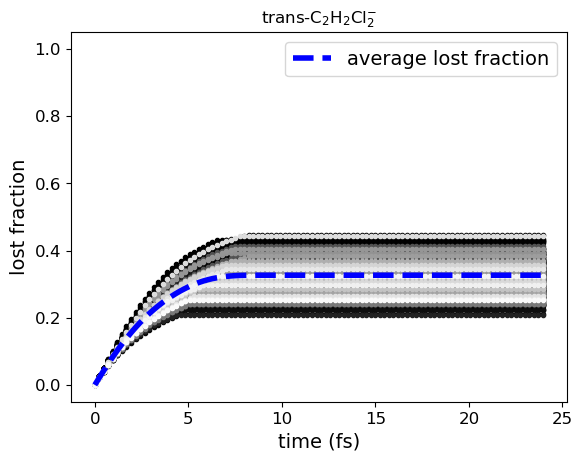} 
\caption{Lost fraction v time for \textit{trans}-dichloroethylene anion.}
\label{fgr:7_lost_fraction}
\end{figure}


\subsection{\textit{1,1}-dichloroethylene, \ch{C2H2Cl2}}
Time step = 10 a.u. (0.2419 fs); 
Number of steps = 1000; 
Temperature = 350 K;
Number of trajectories = 194;
$\eta = 0.00010$ a.u.

Statistics on equilibrium lost fraction: \textit{minimum}= $0.7911$; \textit{maximum}=$0.9990$; 
\textit{mean} =$0.9338$; \textit{median}=$0.9383$; \textit{standard deviation} = $0.0482$.

\begin{figure}[H]
\includegraphics[scale=0.8]{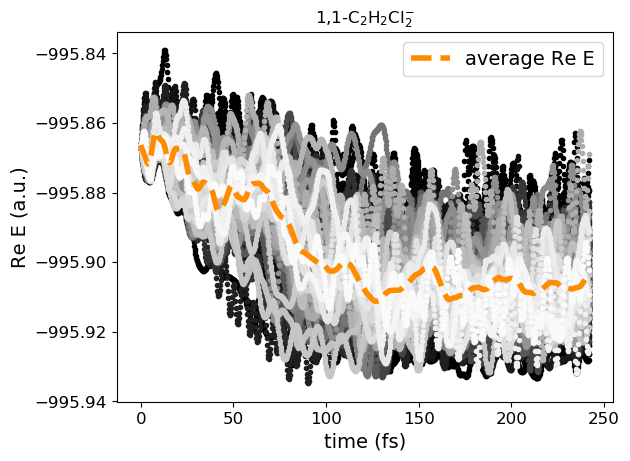} 
\caption{$\Re E$ v time for \textit{cis}-dichloroethylene anion.}
\label{fgr:8_ReE}
\end{figure}

\begin{figure}[H]
\includegraphics[scale=0.8]{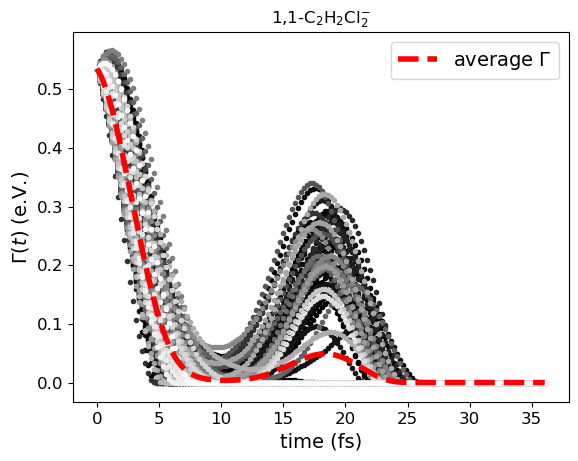} 
\caption{Resonance width, $\Gamma$, v time for \textit{cis}-dichloroethylene anion.}
\label{fgr:8_Gamma}
\end{figure}

\begin{figure}[H]
\includegraphics[scale=0.8]{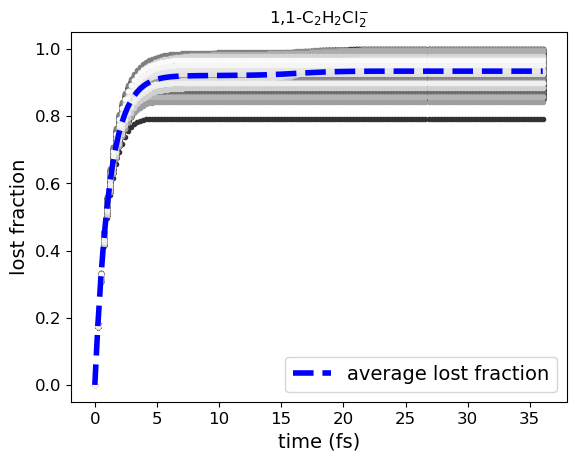} 
\caption{Lost fraction v time for \textit{cis}-dichloroethylene anion.}
\label{fgr:8_lost_fraction}
\end{figure}


\subsection{On the out-of-plane motion of the dichloroethylene anion isomers}
\begin{figure*}
\includegraphics[scale=0.55]{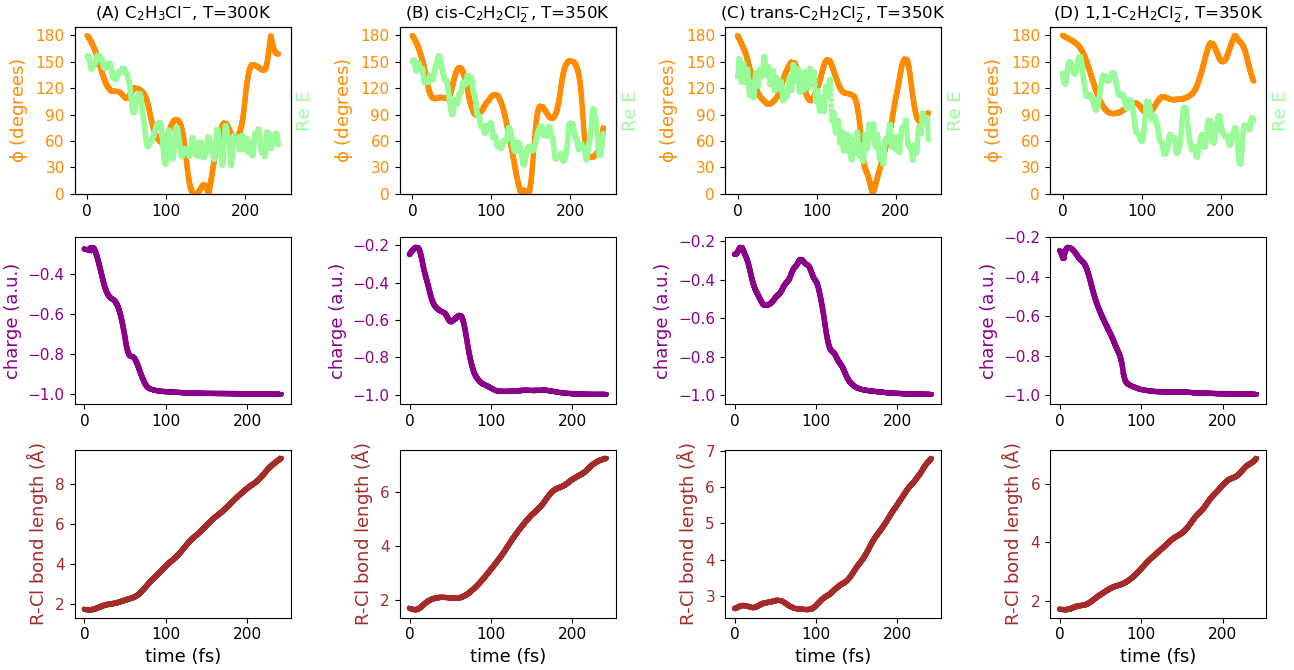} 
\caption{Dihedral angle, Mulliken charge (on dissociating Cl) and R-Cl bond length profiles for the 
trajectories shown in Fig. 3 of article (and Fig. \ref{fgr:chloroethylenes_etc_and_ReE_overlap} above).}
\label{fgr:dihedrals_chloroethylenes_CAP-AIMD}
\end{figure*}

The top row graphs in Fig. \ref{fgr:dihedrals_chloroethylenes_CAP-AIMD} report the dihedral 
angle between the plane containing the two \ch{C} atoms and: i)  the dissociating \ch{Cl}, and 
ii) the other non-C atom geminal to the dissociating \ch{Cl} atom. The normal vectors of these 
two planes are defined such that they are anti-parallel to each other at $t=0$. We observe from 
the dihedral plots that right from the onset of their time evolutions these ethylene derivative TAs 
go through an inexorable decrease in the dihedral angle $\phi$ in question, accompanied by the 
R-\ch{Cl} bond elongation, where \ch{Cl} refers to the dissociating chlorine. In the vast majority of 
our simulations, we see the R-\ch{Cl} bond contracts just a little bit in the beginning before 
elongating as $\phi$ decreases. With the exception of 1,1-dichloroethylene (column D of Fig. \ref{fgr:dihedrals_chloroethylenes_CAP-AIMD}), we also see a wiggly but steep drop of $\phi$ 
towards zero once $90^\circ \lesssim \phi \lesssim 120^\circ$. When $90^\circ \lesssim \phi 
\lesssim 120^\circ$, the R structure is often almost perpendicular to the R-Cl bond; this is 
usually when we begin to see a sharp increase in the R-Cl bond length.

Overall, as shown in the same plots, the sharp decrease in $\phi$ in the first $90-120$ fs follows 
more or less the same trend in time as $\Re E$. The Mulliken charge on the dissociating \ch{Cl} 
also decreases rapidly (see second row in Fig. \ref{fgr:dihedrals_chloroethylenes_CAP-AIMD}) 
in the same interval of time that $\phi$ drops sharply. Interestingly, the time step at which this 
charge begins to stabilize is the same point at which we see a stabilization in $\Re E$.

\subsection{Trichloroethylene, \ch{C2HCl3}}
We ran 24 trajectories, the dissociative electron attachment (DEA) process led to the formation 
of \textit{trans}-dichloroethylene radical + \ch{Cl-} in 21 trajectories, \textit{1,1}-dichloroethylene 
+ \ch{Cl-} in 2 trajectories, and \textit{cis}-dichloroethylene + \ch{Cl-} in 1 trajectory. We show 
below results for the first two channels. Results for the \textit{cis}-dichloroethylene + \ch{Cl-} 
channel not shown below are very similar to the first two.

\subsubsection{\textit{trans}-dichloroethylene channel}
Time step = 20 a.u. (0.4838 fs); 
Number of steps = 1000; 
Temperature = 370 K;
Number of trajectories = 21;
$\eta = 0.00500$ a.u.

Statistics on equilibrium lost fraction: \textit{minimum}= $0.0861$; \textit{maximum}=$0.2229$; 
\textit{mean} =$0.1148$; \textit{median}=$0.1037$; \textit{standard deviation} = $0.0389$.

\begin{figure}[H]
\includegraphics[scale=0.8]{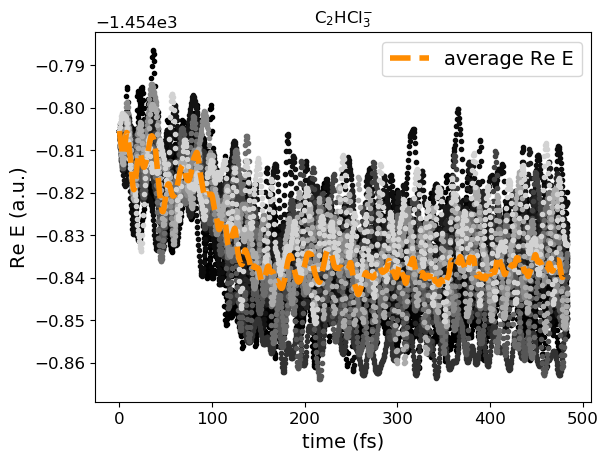} 
\caption{$\Re E$ v time for trichloroethylene anion. An offset of --1454.0 is used for the y-axis.}
\label{fgr:9_transprod_ReE}
\end{figure}

\begin{figure}[H]
\includegraphics[scale=0.8]{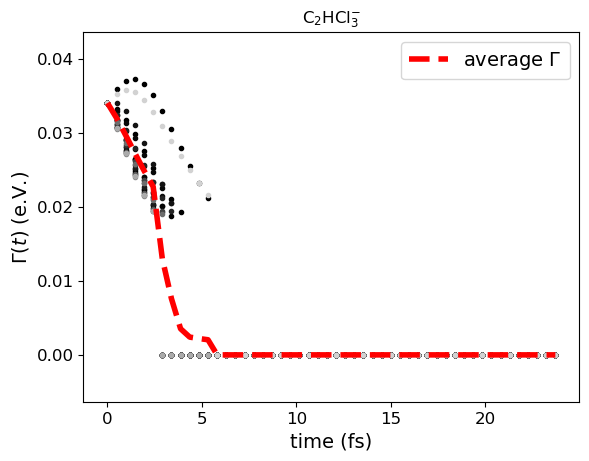} 
\caption{Resonance width, $\Gamma$, v time for trichloroethylene anion.}
\label{fgr:9_transprod_Gamma}
\end{figure}

\begin{figure}[H]
\includegraphics[scale=0.8]{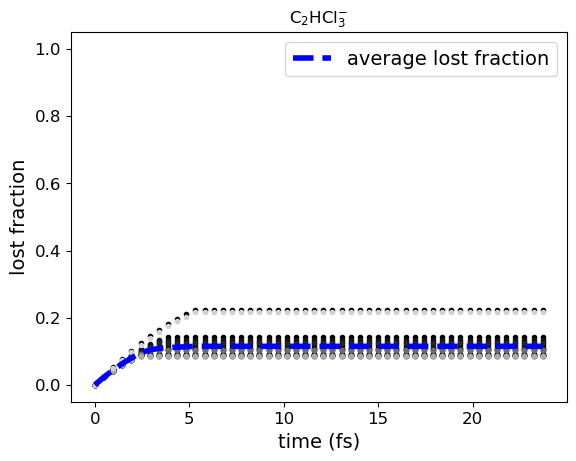} 
\caption{Lost fraction v time for trichloroethylene anion.}
\label{fgr:9_transprod_lost_fraction}
\end{figure}

\subsubsection{\textit{1,1}-dichloroethylene channel}
Time step = 20 a.u. (0.4838 fs); 
Number of steps = 1000; 
Temperature = 370 K;
Number of trajectories = 2; 
$\eta = 0.00500$ a.u.

\begin{figure}[H]
\includegraphics[scale=0.8]{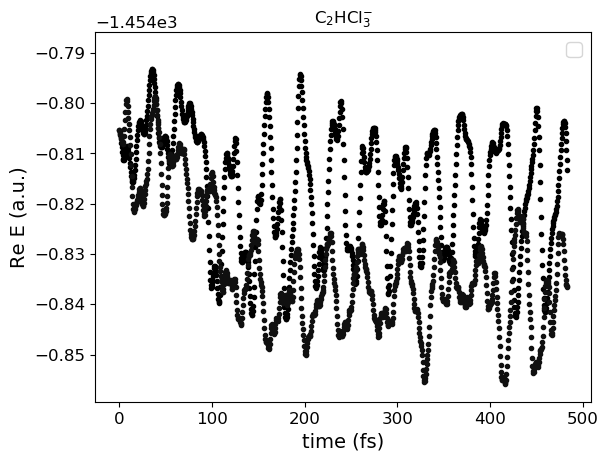} 
\caption{$\Re E$ v time for trichloroethylene. An offset of --1454.0 is used for the y-axis.}
\label{fgr:9_geminiprod_ReE}
\end{figure}

\begin{figure}[H]
\includegraphics[scale=0.8]{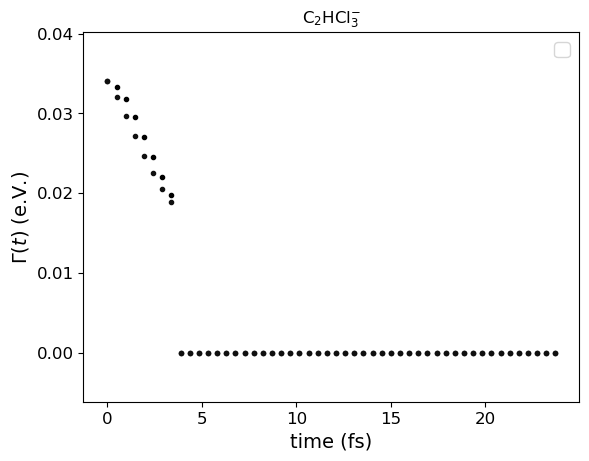} 
\caption{Resonance width, $\Gamma$, v time for trichloroethylene anion.}
\label{fgr:9_geminiprod_Gamma}
\end{figure}

\begin{figure}[H]
\includegraphics[scale=0.8]{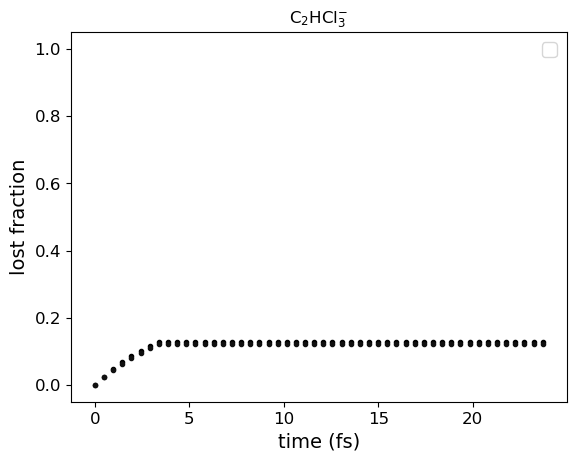} 
\caption{Lost fraction v time for trichloroethylene anion.}
\label{fgr:9_geminiprod_lost_fraction}
\end{figure}


\subsection{Tetrachloroethylene, \ch{C2Cl4}}
Time step = 20 a.u. (0.4838 fs); 
Number of steps = 1000; 
Temperature = 410 K;
Number of trajectories = 21;
$\eta = 0.02340$ a.u.

Statistics on equilibrium lost fraction: \textit{minimum}= $0.0052$; \textit{maximum}=$0.0053$; 
\textit{mean} =$0.0053$; \textit{median}=$0.0053$; \textit{standard deviation} = $3.71 \cdot 10^{-5}$.

The $\Gamma$ profile here, Fig. \ref{fgr:10_Gamma}, shows a very small width for the anion. 
Also, the width becomes zero almost immediately, i.e., after 1 fs. This is the reason why this particular DEA 
can be simulated without CAP. This observation is also in agreement with the very low lost 
fraction that we get from our simulations, see Fig. \ref{fgr:10_lost_fraction}.

\begin{figure}[H]
\includegraphics[scale=0.8]{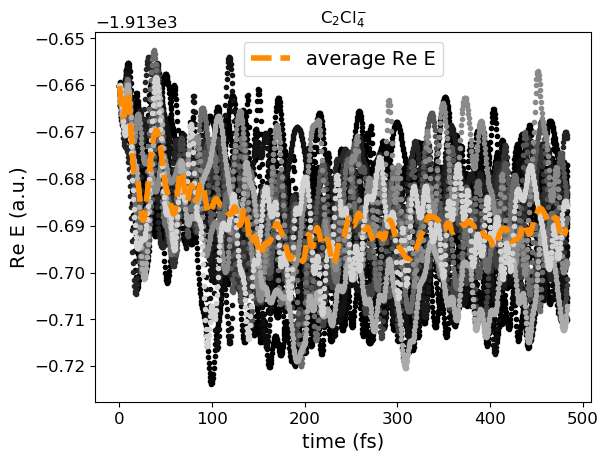} 
\caption{$\Re E$ v time for tetrachloroethylene anion. An offset of --1913.0 is used for the y-axis.}
\label{fgr:10_ReE}
\end{figure}

\begin{figure}[H]
\includegraphics[scale=0.8]{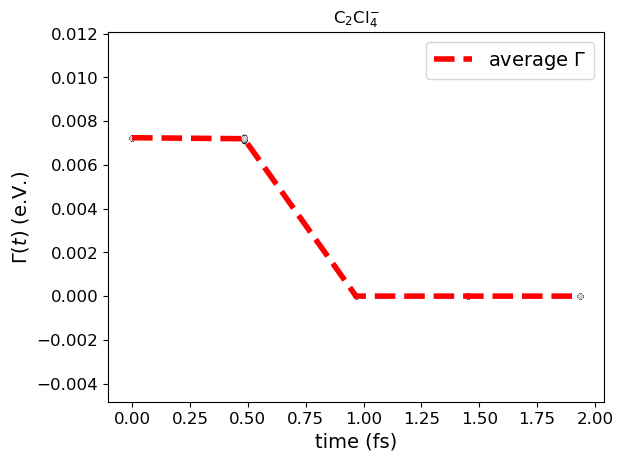} 
\caption{Resonance width, $\Gamma$, v time for tetrachloroethylene anion.}
\label{fgr:10_Gamma}
\end{figure}

\begin{figure}[H]
\includegraphics[scale=0.8]{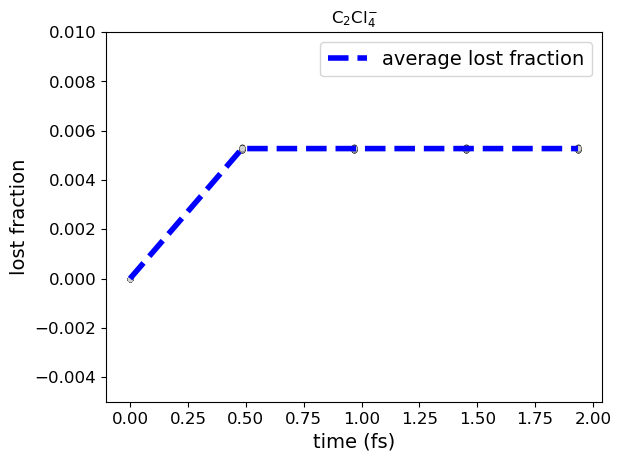} 
\caption{Lost fraction v time for tetrachloroethylene anion.}
\label{fgr:10_lost_fraction}
\end{figure}


\subsection{Chloroethane, \ch{C2H5Cl}}
Time step = 5 a.u. (0.1209 fs); 
Number of steps = 1200; 
Temperature = 260 K;
Number of trajectories = 67.

The resonance width for chloroethane anion is higher compared to the chlorinated ethylene 
anions discussed above because we are dealing with a $\sigma^*$ resonance here. We 
observe a very sharp rise of the lost fraction in the beginning. The average equilibrium lost 
fraction, $\left<f_\infty\right>$, is practically unity, indicating that the DEA process for 
chloroethane anion is very inefficient. This is supported by experimental results \cite{pearl94}.

\begin{figure}[H]
\includegraphics[scale=0.8]{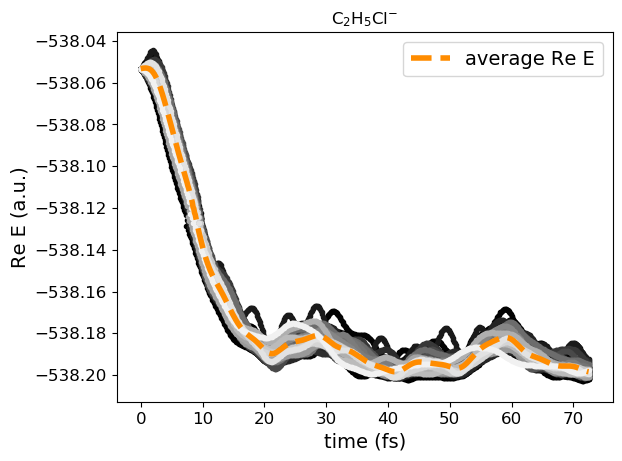} 
\caption{$\Re E$ v time for chloroethane anion.}
\label{fgr:11_ReE}
\end{figure}

\begin{figure}[H]
\includegraphics[scale=0.8]{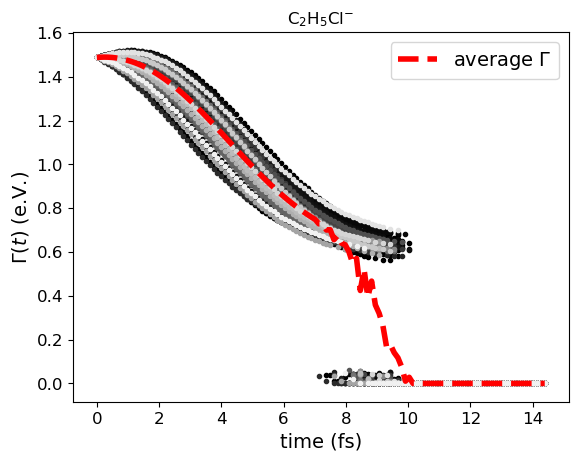} 
\caption{Resonance width, $\Gamma$, v time for chloroethane anion.}
\label{fgr:11_Gamma}
\end{figure}

\begin{figure}[H]
\includegraphics[scale=0.8]{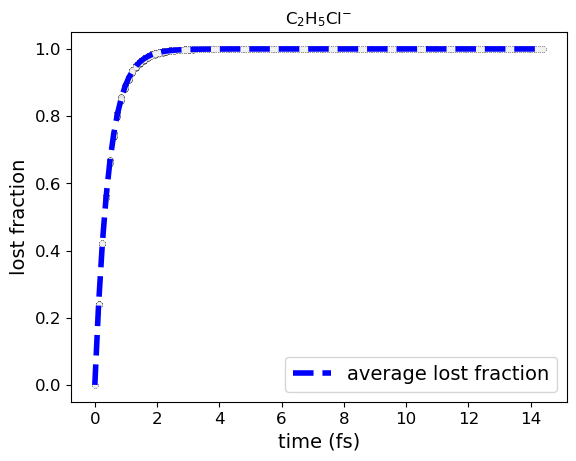} 
\caption{Lost fraction v time for chloroethane anion. The equilibrium lost fraction is $\sim 1$ 
for all trajectories.}
\label{fgr:11_lost_fraction}
\end{figure}

\bibliography{a_CAP_AIMD}